\documentclass[usenatbib,usenatbib]{mnras}
\usepackage{todonotes}
\usepackage{times}
\usepackage{amssymb}
\usepackage{amsmath}
\usepackage{graphicx} 
\usepackage{bmpsize}
\usepackage{epstopdf}
\usepackage{dblfloatfix}    

\newcommand{\be}{\begin{equation}}
\newcommand{\ba}{\begin{eqnarray}}
\newcommand{\ee}{\end{equation}}
\newcommand{\ea}{\end{eqnarray}}

\def\gtsima{$\; \buildrel > \over \sim \;$}
\def\ltsima{$\; \buildrel < \over \sim \;$}
\def\gsim{\lower.5ex\hbox{\gtsima}}
\def\lsim{\lower.5ex\hbox{\ltsima}}
\def\simgt{\lower.5ex\hbox{\gtsima}}
\def\simlt{\lower.5ex\hbox{\ltsima}}
\def\simpr{\lower.5ex\hbox{\prosima}}
\def\la{\lsim}
\def\ga{\gsim}

\def\simless{\mathbin{\lower 3pt\hbox
   {$\rlap{\raise 5pt\hbox{$\char'074$}}\mathchar''7218$}}}  
\def\simgreat{\mathbin{\lower 3pt\hbox
   {$\rlap{\raise 5pt\hbox{$\char'076$}}\mathchar''7218$}}}  
\def\apj{ApJ}
\def\apjs{ApJS}
\def\apjl{ApJL}
\def\aap{A\&A}

\def\nat{Nature}
\def\mnras{MNRAS}

\title[The QSO contribution to the 21-cm signal from the CD]{Evaluating the QSO contribution to the 21-cm signal from the Cosmic Dawn}
\author[H. E. Ross, et al.]
{Hannah E. Ross,$^{1,2,3}$ \thanks{email: HRoss@lbl.gov}
Keri L. Dixon,$^{4,2}$ 
Raghunath Ghara,$^{3}$
Ilian T. Iliev,$^{2}$ and
Garrelt Mellema$^{3}$ 
\\
$^1$ Computational Cosmology Center, Computational Research Division, Lawrence Berkeley National Lab, Berkeley, CA 94720
\\
$^2$ Astronomy Centre, Department of Physics \& Astronomy, Pevensey III Building, University of Sussex, Falmer, Brighton, BN1 9QH,United Kingdom
\\
$^3$ Department of Astronomy \& Oskar Klein Centre, AlbaNova, Stockholm University, SE-106 91 Stockholm, Sweden\\
$^4$ New York University Abu Dhabi, PO Box 129188, Saadiyat Island, Abu Dhabi, UAE\\}
\date{Accepted ?; Received ??; in original form ???}

\pubyear{2019}

\begin{document}
\label{firstpage}
\pagerange{\pageref{firstpage}--\pageref{lastpage}}
\maketitle

\begin{abstract}

The upcoming radio interferometer Square Kilometre Array (SKA) is expected to directly detect the redshifted 21-cm signal from the neutral hydrogen present during the Cosmic Dawn. Temperature fluctuations from X-ray heating of the neutral intergalactic medium can dominate the fluctuations in the 21-cm signal from this time. This heating depends on the abundance, clustering, and properties of the X-ray sources present, which remain highly uncertain. We present a suite of three new large-volume, 349\,Mpc a side, fully numerical radiative transfer simulations including QSO-like sources, extending the work previously presented in Ross et al. (2017). The results show that our QSOs have a modest contribution to the heating budget, yet significantly impact the 21-cm signal. Initially, the power spectrum is boosted on large scales by heating from the biased QSO-like sources, before decreasing on all scales. Fluctuations from images of the 21-cm signal with resolutions corresponding to SKA1-Low at the appropriate redshifts are well above the expected noise for deep integrations, indicating that imaging could be feasible for all the X-ray source models considered. The most notable contribution of the QSOs is a dramatic increase in non-Gaussianity of the signal, as measured by the skewness and kurtosis of the 21-cm probability distribution functions. However, in the case of late Lyman-$\alpha$ saturation, this non-Gaussianity could be dramatically decreased particularly when heating occurs earlier. We conclude that increased non-Gaussianity is a promising signature of rare X-ray sources at this time, provided that Lyman-$\alpha$ saturation occurs before heating dominates the 21-cm signal.

\end{abstract}


\section{INTRODUCTION}
\label{intro}

The Epoch of Reionization, hereafter EoR, is the cosmological era during which the first luminous sources reionized the Universe. Observations of the high-redshift Lyman-$\alpha$ forest \citep[e.g.][]{McGreer2015,Davies2018,Bosman2018}, an observed decrease in Lyman-$\alpha$ emitting galaxies at high redshifts \citep[e.g.][]{Pentericci2014, Tilvi2014,Barros2017,Mason2018}, and temperature measurements of the high-redshift intergalactic medium \citep[IGM; e.g.][]{Raskutti2012, Bolton2012} indicate reionization ended sometime before $z \approx 5.7$. The start of substantial reionization (i.e. more than 10 per cent of the hydrogen mass) is constrained to be redshift 10 by the measured Thomson optical depth for CMB scattering \citep[e.g.][]{Planck2016}. Other than these constraints on the timing of the EoR, the astrophysics of this era remains extremely uncertain.

The most powerful observational probe of this epoch is the redshifted 21-cm signal originating from the hyperfine spin-flip transition of hydrogen. Three experiments are currently attempting to measure the 21-cm signal from the EoR using low-frequency radio interferometry: LOFAR\footnote{\url{http://www.lofar.org/}}, MWA\footnote{\url{http://www.mwatelescope.org/}} and PAPER\footnote{\url{http://eor.berkeley.edu/}}. The LOFAR collaboration has recently set an upper limit to the 21-cm power spectrum from the Epoch of reionization \citep{LOFAR2017} and similar constraints were previously published based on data from GMRT\footnote{\url{http://gmrt.ncra.tifr.res.in/}} \citep{GMRT2011,GMRT2013}. PAPER also placed constraints on the 21-cm power spectrum \citep{Ali2015}, but have since retracted these claims \citep{Ali2018}. The future interferometers HERA\footnote{\url{http://reionization.org/}} and SKA\footnote{\url{https://www.skatelescope.org/}} are expected to be able to detect and possibly image the EoR. 

The beginning of the EoR, when the first luminous sources start to appear but reionization is not yet fully under way, is referred to as the Cosmic Dawn (CD). During this period, the gas temperature fluctuations in the neutral IGM are likely to be the dominant contributor to 21-cm fluctuations (see Section~\ref{sec:theory} for more details). The neutral IGM can only be heated by X-ray photons as they have long mean free paths and are thus able to travel far from their origin, penetrating deep into the neutral hydrogen regions. In contrast, the UV photons produced by stars have short mean free paths and only heat and ionize very locally. Therefore, the 21-cm signal from these early stages of reionization is expected to be sensitive to the spectra, abundance, and clustering of any X-ray sources present at this time. The Experiment to Detect the Global EoR Signature, EDGES\footnote{\url{http://loco.lab.asu.edu/edges/}}, has recently claimed to have detected an extremely strong absorption signal from the CD \citep{Bowman2018}. If this result is confirmed then additional physics is required to explain the measured signal. The observational difficulties of this experiment and concerns over the foreground modeling \citep{Hill2018} lead us to conclude that further validation from another independent observation is required in order to verify the result. 

The nature of X-ray sources present in the CD sources remains extremely uncertain. Simulations have suggested that the first generation of stars (Population III stars, referred to as Pop III stars hereafter) could have formed binary systems as early as redshift 30 \citep{Glover2003}. High-mass X-ray binaries, HMXBs, have therefore been suggested as significant contributors to the X-ray emissions \citep[e.g.][]{Xu2014,Jeon2014,Jeon2015}. QSOs are also a likely candidate for early X-ray heating. \citet{Chardin2015} argued that the observation of a large scatter in the Lyman-$\alpha$ opacity in \citet{Becker2015} suggests patchy hydrogen reionization, implying the presence of rare, bright sources. The presence of high-redshift QSOs is also consistent with the gentle slope at the bright-end of the high-redshift UV galaxy luminosity at $z\sim$7 \citep{Bowler2012,Bowler2014,Bowler2015} and the X-ray spectra associated with these galaxies \citep{Stark2015a,Stark2015b, Stark2017,Mainali2017}. An observation of high-redshift, low-luminosity QSOs in \citet{Giallongo2015} has suggested that the low-mass end of the QSO X-ray luminosity functions (QXLF) may be steeper than previously thought. \citet{Grissom2014, Giallongo2015, Chardin2015, Khaire2016} and \citet{Mitra2016} argue that QSOs may even be numerous enough to contribute significantly to reionization itself. Contrarily, \citet[e.g.][]{Onoue2017,Onorbe2017,Qin2017,Hassan2018,Akiyama2018} and \citet{Parsa2018} argue that QSOs are unlikely to contribute significantly to reionization.

Multiple theoretical studies have previously investigated the impact of X-ray heating during the CD. Early work on this topic was analytical \citep[e.g.][]{Glover2003,Furlanetto2004} and considered a simple X-ray background. However, most recent works have focused on semi-numerical \citep[see e.g.][]{Santos2010,Mesinger2013, Fialkov2014,Knevitt2014,Ghara2016,Ghara2017,Greig2018,Das2017,Douna2018} and numerical \citep[eg.][]{Xu2014,Ahn2015,Ross2017} modeling. These works consider the dominant contributors of X-rays in the CD to be HXMBs and, therefore, trace the stellar population. There is disagreement on the contribution of HMXBs. For example, \citet{Knevitt2014} find that HMXBs do not contribute significantly to the CD; whereas, \citet{Greig2018} predict that the heating during the CD will be detectable by SKA. These differences stem from the lack of understanding of high redshift HMXBs, and of high redshift sources in general.

QSO source models have also been investigated in previous works both using semi-numerical \citep[e.g.][]{Yajima2014,Datta2016,Qin2017,Hassan2018} and numerical \citep[e.g.][]{Baek2010,Kakiichi2017,Semelin2017,Eide2018} methods. The large scale fully numerical simulations run by \citet{Kakiichi2017} and semi-numerical simulations run by \citet{Hassan2018} and \citet{Qin2017} focus on the impact of QSOs during the EoR rather than the CD. \citet{Datta2016} investigate an individual QSO and assume the luminosity to be the same as a the high redshift QSO observation from \citet{Mortlock2011}. They focus primarily on the detectability of an individual QSO source rather than their contribution to reionization and heating process.

The luminosities of the QSOs are often calculated by assuming black hole masses are proportional the mass of stars in the galaxy and are accreting at the Eddington limit. For example, \citet{Yajima2014} have considered individual sources in a comparable way to \citet{Datta2016}, but assuming the luminosity of QSOs to be proportional to the mass of the halo. While there has been an observed correlation between the mass of the bulge of galaxies and the masses of their central black holes \citep[e.g.][]{2004Haring,Lasker2016}, it has been known for a long time that this relationship does not extend to the luminosity of the QSO \citep[e.g.][]{Woo2002}. \citet{Baek2010} and \citet{Eide2018} have used this assumption that $L \propto M_{\rm halo}$ in their detailed fully numerical simulations of QSOs during the CD including the radiative transfer ionizing UV photons, X-rays, and Lyman-$\alpha$ photons, albeit with a smaller boxsize ($\sim$ 100~Mpc $h^{-1}$). 

In this paper, we extend our suite of numerical simulations of the inhomogeneous heating during the CD previously presented in \cite{Ross2017}, hereafter referred to as Paper I, with the addition of different X-ray emitter models. Using the same cosmic density fields and halo lists, we compare the morphology and evolution of the 21-cm signal for these different cases. We also include a Lyman-$\alpha$ background in order to comment on the possible impact of late Lyman-$\alpha$ saturation. Our simulations are sufficiently large to capture the large-scale patchiness of reionization and to make statistically meaningful predictions for future 21-cm observations.

The outline of the paper is as follows. In Section~\ref{sec:method}, we present our $N$-body and radiative transfer simulations and describe our different X-ray source models. In Section~\ref{sec:theory}, we summarize the extraction of the 21-cm signatures from our simulations. In Section~\ref{sec:lya}, we describe our semi-numerical radiative transfer of Lyman-$\alpha$ photons. Section~\ref{sec:results} contains our results, primarily comparisons between the different source models. We then conclude in Section~\ref{sec:conclusions}. The cosmological parameters we use throughout this work are ($\Omega_\Lambda$, $\Omega_\mathrm{M}$, $\Omega_\mathrm{b}$, $n$, $\sigma_\mathrm{8}$, $h$) = (0.73, 0.27, 0.044, 0.96, 0.8, 0.7); where the notation has the usual meaning and $h = \mathrm{H_0} / (100 \  \mathrm{km} \ \rm{s}^{-1} \ \mathrm{Mpc}^{-1}) $. These values are consistent with the latest results from WMAP \citep{Komatsu2011} and Plank combined with all other available constraints \citep{Planck2015,Planck2016}.

\section{METHODOLOGY}
\label{sec:method}

\subsection{THE SIMULATIONS}
\label{sec:sims}

The density fields and halo catalogues were obtained from a high-resolution, $N$-body simulation (presented in \citet{2016MNRAS.456.3011D}, and also used in Paper I). This simulation was run using the \textsc{\small CubeP$^3$M} Code \citep{Harnois2013} and followed $4000^3$ particles in a $349\,$~comoving~Mpc per side volume to enable reliable halo identification down to $10^9\, \rm M_\odot$. 

The radiative transfer simulations were run using the latest version of the photon conserving, short characteristics ray-tracing method \textsc{\small C$^2$-Ray} code \citep{Mellema2006} over a grid-size of 250$^3$ with a timstep of 11.52~Myrs. The original version was updated in order to accommodate higher energy photons and track the temperature of the IGM \citep{Friedrich2012}. These modifications included the inclusion of multi-frequency heating, the inclusion of all three species of helium, and a full on-the-spot approximation in order to model secondary ionizations. To be able to correctly handle unresolved ionized regions, we developed a new, multiphase, approach, described below in Section~\ref{sec:multiphase} and Appendix~A.

A total of five simulations are presented: a simulation with only stellar sources (S1), another two with both stellar sources and QSO sources with power laws of different spectral indices (S2, S3), a simulation with both stellar and HMXB sources (S4), and a simulation with stellar sources, HMXBs, and the harder QSO sources (S5). Information about the spectral indices of the X-ray sources are given in Table 1, and the source details are outlined in Section~\ref{sources}. S1 and S4 have previously been presented in Paper I and S5 in \citet{Ross2018}. S2 and S3 were run with the new multiphase version of the code.

\subsection{SOURCES}
\label{sources}

\begin{figure} 
\centering
\includegraphics[width=1.0\columnwidth]{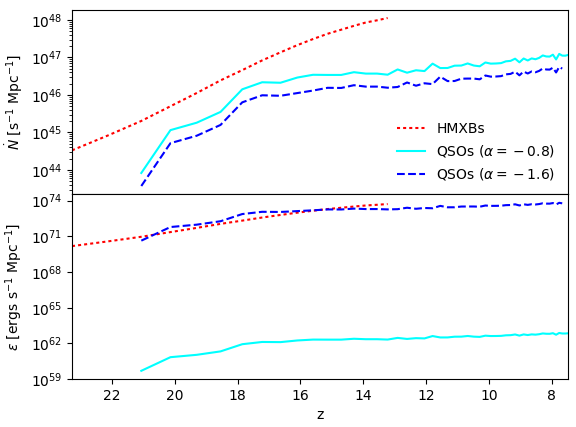}
\caption{The ionizing photon rate (top panel) and emissivity (bottom panel) of each source type. $f_{\rm X}$ is not fundamental to our method so we do not show it here.}
\label{fig:em}
\end{figure}

We consider three types of ionization sources: stellar sources, HMXBs, and QSOs. The total number of ionizing photons, $\dot{N}$, and emissivity, $\epsilon$, of each type of source are shown in Fig.~\ref{fig:em}. We plot $\dot{N}$ and $\epsilon$ rather than $f_{\rm X}$ as the calculation of the later requires assumptions about $N_{\rm i}$ and $f_{\rm esc}.$

\subsubsection{Stellar sources}

Stellar sources form inside dark matter haloes. High-Mass Atomically Cooling Haloes, HMACHs, are haloes with masses greater than $10^9 M_\odot$ and are resolved by the $N$-body simulation, from which they were extracted using the spherical over-density algorithm. HMACHs have sufficiently deep gravitational wells to accrete surrounding IGM gas, even when the gas has been photo-heated. Furthermore, their gas virial temperatures are above $10^4$~K, allowing the gas to efficiently cool through hydrogen and helium atomic line radiation. Therefore, these haloes are able to keep accreting fresh gas and forming stars regardless of their local ionization. We therefore assume these sources to be un-suppressible, i.e. continually forming stars end emitting photons into the IGM.

Haloes of lower mass are not able to accrete IGM gas with temperatures around $10^4$~K; however, haloes above $10^8 M_\odot$ are capable of accreting cold IGM gas, as it can cool to sufficiently low temperatures through atomic lines. These Low Mass Atomically Cooling Haloes, LMACHs, are able to gravitationally bind neutral, but not ionized gas. In reality there is no sharp cut-off between HMACHS and LMACHs, rather a gradual decrease in the amount of cold gas which can be retained with declining halo mass \citep{Efstathiou1992,Navarro1997,Dijkstra2004,Hasegawa2013}. However, for simplicity we assumed that LMACHs residing in cells that are more than 10 percent ionized are assumed to not form stars, i.e. they are suppressed \citep{2007MNRAS.376..534I,2016MNRAS.456.3011D}. This is in rough agreement with recent results from detailed fully-coupled radiative hydrodynamics simulations, though the suppression mass and level of suppression depend on the strength of supernova feedback \citep{Wise2008,2016MNRAS.463.1462O,2018MNRAS.480.1740D,2018arXiv181111192O}.

LMACHs are not resolved by the $N$-body simulation and are added using a sub-grid model \citep{Ahn2015a} that is calibrated against higher mass resolution simulations of smaller volumes. We do not take into account the redshift dependence of the cut-off mass at which haloes are no longer able to accrete ionized gas; however, this effect is minor \citep{1994ApJ...427...25S}.

Stellar sources are assigned a blackbody spectrum with an effective temperature of $T_{\mathrm{eff}} = 5 \times 10^4$~K, which is consistent with observations of O and B stellar spectra. The luminosity of the sources is proportional to the mass of their host haloes:
\begin{equation}
\dot{N_\gamma} = g_\gamma \frac{M \Omega_{\rm b}}{m_{\rm p} (10~\rm Myr) \Omega_0},
\label{eq:norm}
\end{equation}

\noindent where $g_\gamma$ is the photon production efficiency factor. Here, $g_\gamma$ is given by: 
\begin{equation}
g_\gamma = f_* f_{\rm esc} N_{\rm i} \frac{10~\rm Myr}{\Delta t},
\label{eq:ggamma}
\end{equation}
where $f_*$ is the star formation efficiency, $f_{\rm esc}$ is the escape fraction, $N_{\rm i}$ is the ionizing photon efficiency per stellar baryon, and $\Delta t$ is the lifetime of the source. LMACHs are given a higher value of $g_\gamma$ (7.1) than HMACHs (1.7), which reflects the likely presence of larger, more efficient Pop~III stars \citep{Stacy2016} and/or higher escape fractions \citep[e.g.][]{Xu2016,Chisholm2018}. These stellar spectra produce minimal X-rays, so do not contribute significantly to the heating of the neutral IGM. For more details on the implementation of these sources, see LB2 of \citet{2016MNRAS.456.3011D}.

\subsubsection{HMXBs}

The HMXB sources are assumed to reside in dark matter haloes and to have power-law spectra:
\begin{equation}
L_{\rm h}(\nu) \propto \nu^{-\alpha^{\rm h}_{\rm x}},
\end{equation}
where $\alpha^{\rm h}_{\rm x} = 1.5$. The energy range extends from 272.08 eV to 100 times the second ionization of helium (5441.60 eV). The low-frequency cut-off corresponds to the obscuration suggested to be present by observational works \citep[e.g.][]{Lutivnov2005} and is consistent with the optical depth from high-redshift, gamma-ray bursts \citep{Totani2006,Greiner2009}. As with the stellar sources, the luminosity is related to the mass of the host halo via equation~(\ref{eq:norm}), but with a lower value of $g_\gamma$ (0.086) for all haloes. $g_\gamma$ is given by equation~(\ref{eq:ggamma}) with $f_{\rm esc}$=1 and $N{\rm _x}$ (the number of X-ray photons produced per stellar baryon) in place of $N{\rm _i}$.  The X-ray luminosities are roughly consistent with measurements of X-ray binaries in local, star-bursting galaxies \citep{Mineo2012}. For more details on the implementation of these sources, see Paper I.
 
\subsubsection{QSO sources}

Due to the lack of suitable high redshift observations, we must make some assumptions about the luminosity and spectra of the QSOs present during the CD. We assume that our QSO-like sources only produce X-rays, which physically means that stellar sources dominate the lower frequency photon budget.\footnote{Note that by considering only the X-ray photons, we are neglecting the UV contribution from QSOs. Within our model and at these high redshifts, we do not expect a large impact on our results, but some comparisons to other studies may complicated by this fact.} The X-ray emissivity from QSOs is quantified using the QXLF. Here, we follow \citet[][see section~6.2 for the functional form and table~4 for the parameter values]{Ueda2014}, though modified to reflect the uncertainty in the higher redshift behaviour. In particular, we alter the high-redshift (in this case, $z>3$) density evolution parameter (known as $p3 = -6.2$ in \citealt{Ueda2014}) to a smaller -2, which is more in line with \citet{Giallongo2015} or generally including more QSOs at high redshift. Furthermore, one aim of this paper is to investigate the maximal impact of these type of sources; though, we concede that such a shallow QSO density evolution is unlikely at the highest redshifts. This QXLF takes the form of a double power law with luminosity-dependent density evolution and is taken over an X-ray luminosity ($L_{\rm X}$) range of $10^{42} - 10^{47}$ ergs s$^{-1})$. 

The number density of QSOs in our simulation volume, $n_{\rm q}$, is calculated by integrating the QXLF, $\Phi(L,z)$, for each redshift
\begin{equation}
n_{\rm q}(z) = \int\limits_{L_{\rm min}}^{L_{\rm max}} \Phi (L,z) \  dL,
\end{equation}
where $\Phi \propto (1+z)^{-2}$ is the QXLF. At high redshift, the number density of haloes capable of hosting quasars (i.e., the HMACHs number density) is insufficient to replicate the QXLF (see below for host halo details).

The QSO spectrum is assumed to be:
\begin{equation}
L_{\rm q}(\nu) \propto \nu^{-\alpha^{\rm q}_{\rm x}},
\end{equation} 
where $\alpha^{\rm q}_{\rm x}$ = 0.8 \citep{Ueda2014} or 1.6 \citep{Brightman2013} for our two QSO models. More generally, we chose a hard and soft model, where the exact spectral indices are unimportant, for comparison sake. The luminosity of each QSO at 2~keV is then assigned by randomly sampling the QXLF. Given this luminosity and spectral index, $\dot{N_\gamma}$ is calculated for the same frequency (energy) range as the HMXBs.

The luminosities of observed QSOs do not correlate with the mass of their host galaxy or that of the central black hole, but rather depend on the physics of the accretion disk \citep[e.g.][]{Woo2002,Middei2017}. Consequently, in our simulations, we place the active QSOs in random HMACH haloes.\footnote{In rare cases, an HMACH will disappear for some reason, such as stripping or merging. In this circumstance, we reassign a brand new QSO elsewhere that is sampled from the current redshift's QXLF.} Initially, HMACHs are too rare to host sufficient numbers of QSOs to reproduce the (admittedly optimistic) luminosity function. At these early times, we assign a single QSO to each existing HMACH halo. Furthermore, in our simulations, we assume a QSO lifetime of 34.56~Myr, which is consistent with current estimates \citep[e.g.][]{Borisova2016,Khrykin2016}. It has been long known that the luminosity of QSOs varies with time over all frequencies, including the X-ray range of the spectrum \citep[e.g.][]{Halpern1984,Pan1990,Jin2017}, and we mimic this behaviour by assigning a new $L_{\rm q}$ every 11.52~Myr, with a value that is within an order of magnitude of the previous $L_{\rm q}$.

\section{The 21-CM SIGNAL}
\label{sec:theory}

We are primarily interested in the 21-cm signal observable, the differential brightness temperature with respect to the CMB, $\delta T_{\mathrm{b}}$, given by:
\begin{equation}
\delta T_{\mathrm{b}} = \left(1 - \frac{T_{\mathrm{CMB}}}{T_{\mathrm{S}}}\right) \ \frac{3 \lambda_0^3 A_{10} T_\star n_{\mathrm{HI}}(z)}{32 \pi H(z) (1+z)}, 
\label{eq:dif}
\end{equation}
where $\lambda_0=21.1$~cm is the line rest-frame wavelength, $A_{10}=2.85\times10^{-15}\,\rm s^{-1}$ is the Einstein $A$-coefficient for spontaneous emission from the triplet to singlet state, $n_{\mathrm{HI}}$ is the density of neutral hydrogen, and $T_{\rm S}$ is the spin temperature.

The $T_{\rm S}$ reflects the relative number of atoms in the singlet and triplet state of the 21-cm line, given by \citep{Field1958}:
\begin{equation}
T_{\rm S} = \frac{T_{\rm CMB}+y_\alpha T_\alpha + y_{\rm c} T_{\rm K}}{1+y_\alpha+y_{\rm c}},
\end{equation}
where $T_\alpha$ is the radiation colour temperature, $T_{\rm K}$ is the kinetic temperature of the gas, and $y_\alpha$ and $y_{\rm c}$ are the coupling coefficients corresponding to the Lyman-$\alpha$ decoupling (the Wouthuysen-Field effect) and collisional decoupling, respectively.  
In the absence of decoupling mechanisms (collisions between atoms or Lyman-$\alpha$ pumping), the 21-cm line is in equilibrium with the CMB, and thus $T_{\rm S}=T_{\rm CMB}$. 

The $y_\alpha$ is calculated using:
\begin{equation}
y_\alpha = \frac{T_*}{T_\mathrm{K}} \frac{P_{10}}{A_{10}},
\end{equation}
where $P_{10}$ is the radiative de-excitation rate due to Lyman-$\alpha$ photons ($\sim 10^9 J_\alpha$, where $J_\alpha$ is the Lyman-$\alpha$ flux), and $T_*=0.068$~K is the temperature corresponding to the energy difference between two the states. 

In the case of early Lyman-$\alpha$ saturation, $y_\alpha$ becomes large everywhere, and $T_{\rm S} \approx T_{\rm K}$, as we assumed in Paper I. However, in the case of late Lyman-$\alpha$ saturation, we must include the additional fluctuations as a post-processing step (see Section~\ref{sec:lya}). 

It is not necessary to include collisional coupling at the redshifts we consider. The Universe has expanded enough that the density of the IGM is insufficient to produce non-negligible collisions between hydrogen atoms. In addition, we do not resolve the dark matter filaments with sufficient densities for collisional coupling to become efficient on the RT grid.

\begin{table} 
\begin{center}
\caption {\label{tab:runs} Table showing the spectral index of our X-ray sources in different runs. All simulations include stellar sources with a blackbody spectrum corresponding to a temperature of 5 $\times 10^4$ K.}
\begin{tabular}{ | l | c | c | c | c | r |}
\hline
Spectral index \ \ \ \ \ & \ \ \ \ S1 \ \ \ \ & \ \ \ \ S2 \ \ \ \ & \ \ \ \ S3 \ \ \ \ & \ \ \ \ S4 \ \ \ \ & \ \ \ \ S5 \ \ \ \  \\
\hline \hline
\ \ \ \ \ \ \ \ \ $\alpha^{\rm h}_{\rm x}$ & -- & -- & -- & 1.5 & 1.5 \ \ \ \ \\ 
\hline
\ \ \ \ \ \ \ \ \ $\alpha^{\rm q}_{\rm x}$ & -- & 0.8 & 1.6 & -- & 0.8 \ \ \ \ \\
\hline
\end{tabular}
\end{center}
\end{table}

\subsection{$T_{\rm K}$ AND RESOLUTION EFFECTS}
\label{sec:multiphase}

It is not computationally feasible to resolve the small scales (of the order of kpc) corresponding to the width of ionization fronts or (more importantly in the CD) small H~II regions while including the larger scales. In the original version of \textsc{\small C$^2$-Ray}, cells containing ionization fronts or small H~II regions are partially ionized and contain an averaged temperature and ionized fraction. While these average values themselves are correct, using them to calculate $\delta T_{\rm b}$ can yield an incorrect result. Consider a cell that contains a small, hot, ionized bubble and a cold, neutral region.  When calculating $\delta T_{\rm b}$, the ionized region should not contribute, but in this case the high temperature of the ionized region will dramatically increase the average value the $T_{\rm K}$ of hydrogen in the cell, and hence the $\delta T_{\rm b}$, will be overestimated. This problem is particularly pronounced during the CD when many small ionized regions enclosed in single cells are present due to our mass resolution being much higher than our RT resolution.

As described in detail in Paper~I, this problem was resolved via post-processing of the simulation using a second simulation without X-ray sources. While sufficient at the redshifts and for the X-ray source model of the simulations in Paper I, this method breaks down at $z \sim 11$, as a significant number of stellar sources are able raise the temperature high enough for collisional cooling to become efficient. More efficient cooling in the X-ray simulation compared to the stellar simulation results in the temperature being underestimated (see Appendix~\ref{sec:limits} for more details on this). 

For the current work, we have developed a new version of \textsc{\small{C$^2$-Ray}} that includes a multiphase, subgrid modeling of the IGM to track the temperatures of the ionized and neutral IGM separately. Not only is this method much more robust and accurate, but it is also significantly more computationally efficient as only one simulation is required. Appendix A provides a more detailed description of our new algorithm and test boxes, and Appendix B presents results of two tests that demonstrate the differences between the old and new algorithm.

\section{LYMAN-$\alpha$ coupling}
\label{sec:lya}

The amount of emitted soft (i.e. non-ionizing) UV photons that subsequently redshift into the Lyman-$\alpha$ resonance is very uncertain. Here we consider the two extremes: very early and very late Lyman-$\alpha$ saturation scenarios in order to demonstrate the full range of these uncertain source parameters. We leave a more detailed analysis of the impact of Lyman-$\alpha$ photons for future work. 

Early mini-haloes (haloes with masses below $\sim 10^8$M$_\odot$ that host the First Stars) may contribute significantly to the Lyman-$\alpha$ background, which could result in Lyman-$\alpha$ saturation occurring quite early. In the most extreme scenario, a strong Lyman-$\alpha$ background may have already been built up by mini-haloes before our simulation has begun ($z\sim23$), as was assumed in Paper I.

At the other end of the possible range, mini-haloes might contribute very little Lyman-$\alpha$ radiation, so the Lyman-$\alpha$ background is only built up by the HMACHs and LMACHs. In this scenario, we must consider the inhomogeneous background they produce. 

Fully numerical radiative transfer simulations of Lyman-$\alpha$ are computationally expensive and, in this case, largely unnecessary. On large scales, a $1/r^2$ profile  (where $r$ is the radial distance from the source) has been shown to be consistent with detailed radiative transfer in \citet{Semelin2007,Vonlanthen2011} and \cite{Higgins2012}. In addition, the Lyman-$\alpha$ photons from a point source have been shown to produce a nearly spherical profile even in the presence of density fluctuations \citep{Vonlanthen2011}.  

Here we employ the method used previously in \cite{Ghara2016}, assuming this $1/r^2$ spherical profile with a few improvements. In order to calculate the Lyman-$\alpha$ flux, $J_\alpha$, from the simulations, we use SEDs generated by the stellar population synthesis code  \textsc{\small{PEGASE2}} \citep{Fioc1999}. Galaxies are initially metal-poor (10$^{-3}$~Z$_\odot$, where Z$_\odot$ is solar metallicity) and are assumed to have a Salpeter initial mass function (with stars with masses between 1 and 100 M$_\odot$). Our method was updated compared to previous versions so that the flux of the Lyman-$\alpha$ photons depend on the mass of the source when they were emitted (i.e. to use the retarded luminosity). This allows us to take into account the variations in luminosity of the sources due to the changing mass of dark matter haloes, the suppression of LMACHs and the movement of the haloes. 

For these calculations, we assume the escape fraction of the ionizing photons to be 10 per cent, and the escape fraction of non-ionizing UV photons to be 100 per cent. Some observations have shown high redshift galaxies to be dustier than expected \citep[e.g.][]{Vieira2013,Watson2015,Laporte2017,Chiaki2019,Tamura2019}. It is important to note that it is possible that early dust enrichment could reduce the fraction of Lyman-$\alpha$ photons that escape.

\begin{figure} 
\centering
\includegraphics[width=1.0\columnwidth]{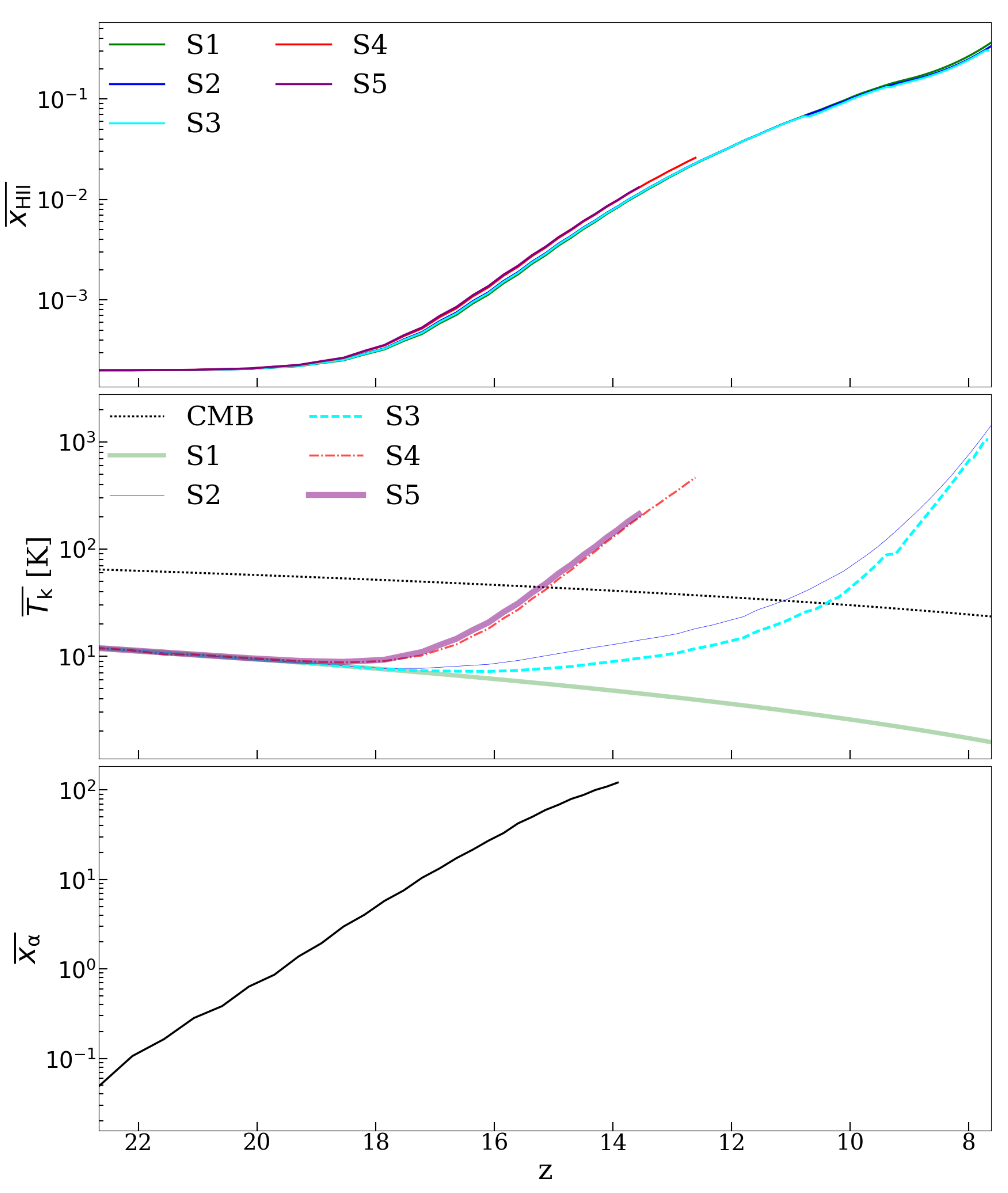}
\caption{The global histories of the volume-averaged ionized fraction ($\overline{x_{\rm H}}$; top), the volume-averaged kinetic temperature of the neutral gas ($\overline{T_{\rm K}}$; middle) and the volume-average value of the Lyman-$\alpha$ coupling coefficient ($\overline{x_{\rm \alpha}}$; bottom).
Note that some curves have significant overlap, particularly for the reionization history.}
\label{fig:global}
\end{figure}

\begin{figure*}
\centering
\includegraphics[width=1.\textwidth]{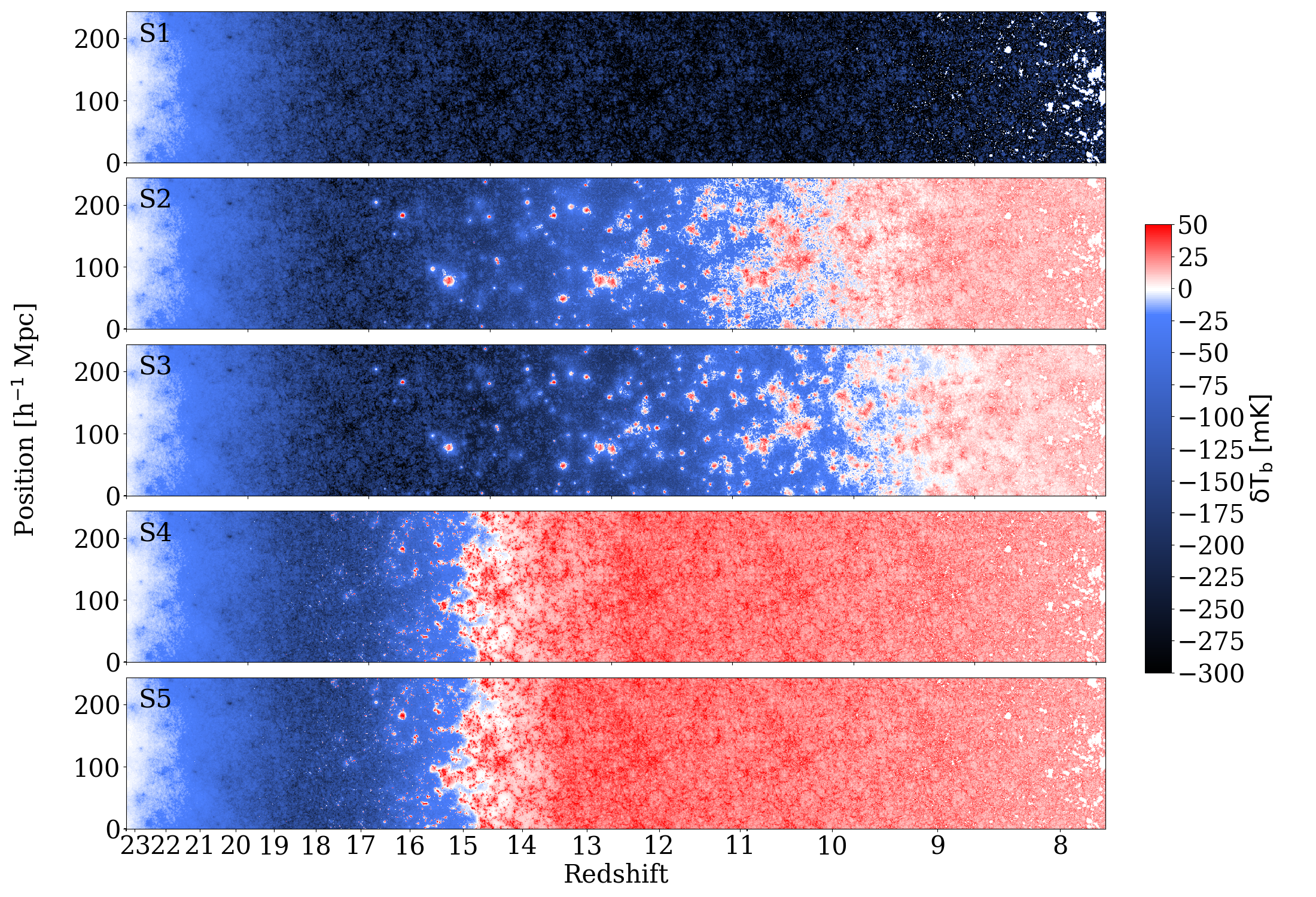}
\caption{The position-redshift lightcone of $\delta T_{\rm b}$ from the five different simulations considered in this study. Here, we can see the different geometries, evolutions, and timings produced by the different source models. The details of the source models are given in Table~\ref{tab:runs}. These lightcones are for the case of late Lyman-$\alpha$ saturation.}
\label{fig:lightcones:dbt}
\end{figure*}

\begin{figure*}
\centering
\includegraphics[width=1.\textwidth]{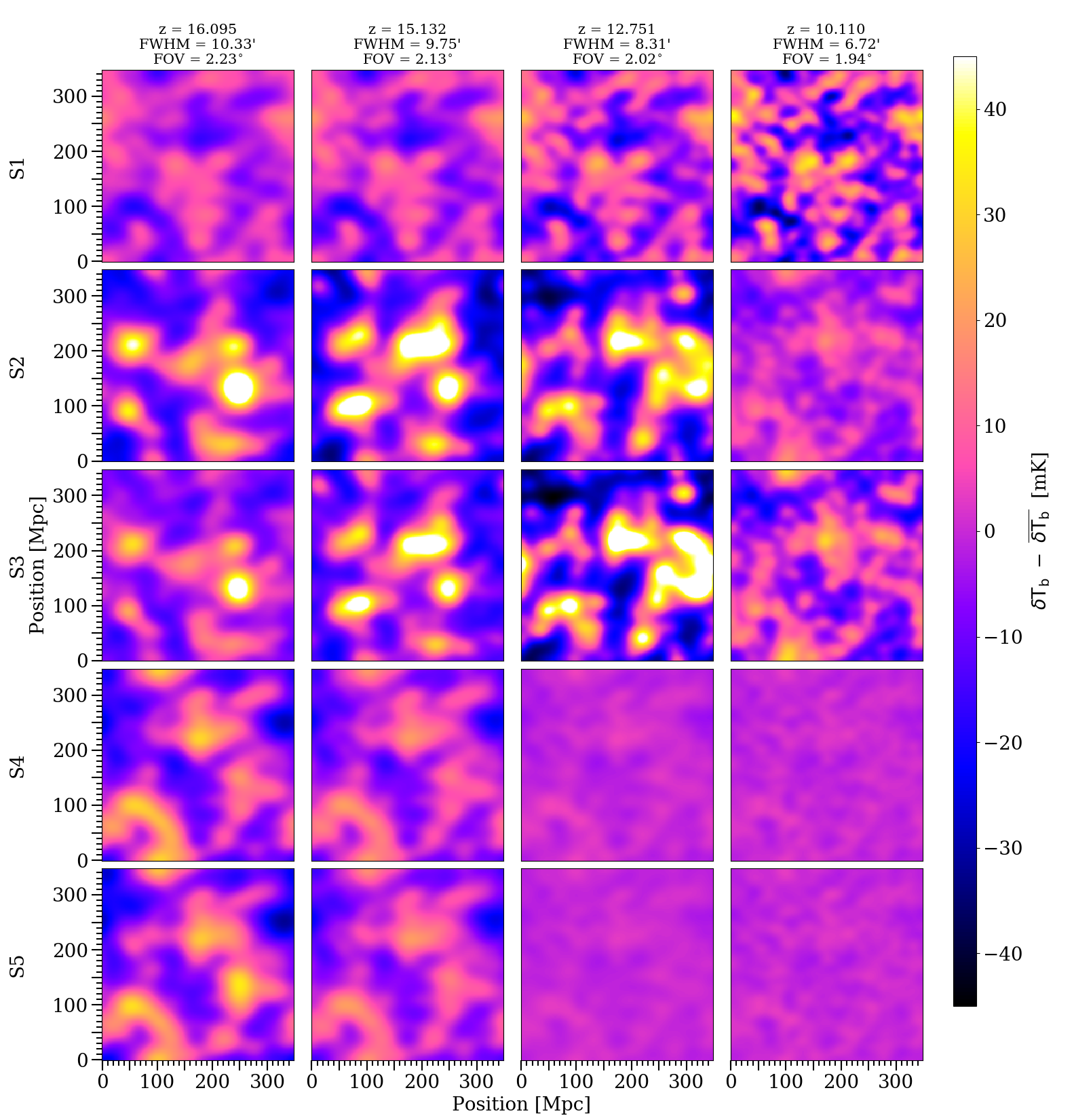}
\caption{Mean-subtracted $\delta T_{\rm b}$ maps, smoothed with a Gaussian beam with a FWHM corresponding to a 1.2~km maximum baseline at the relevant frequency and bandwidth-smoothed with a top hat function (width equal to the distance corresponding to the beam width). The rows correspond to our five models. The columns represent higher to lower redshift from left to right.}
\label{fig:blurred}
\end{figure*}

\section{RESULTS}
\label{sec:results}

\subsection{Global Histories}

In Fig.~\ref{fig:global} we show the global histories of the ionization, neutral gas temperature and Lyman-$\alpha$ coupling coefficient of our simulations. The top panel shows the volume averaged ionized fraction, $\overline{x_{\rm H}}$. All sources follow a similar reionization history as, despite producing photons capable of ionizing more than one atom, X-ray sources yield far fewer photons overall than their stellar companions.  

Although X-ray photons are rare, they lead to significant photo-heating as we can see from the middle panel of Fig.~\ref{fig:global} which displays the volume averaged temperature, $\overline{T_{\rm K}}$. X-ray heating progresses more rapidly in the cases including HMXBs than those with only QSOs.
 
Finally, in the bottom panel the volume averaged value of the Lyman-$\alpha$ coupling coefficient, $\overline{x_{\rm \alpha}}$, is shown. $\overline{x_{\rm \alpha}}$ follows the same evolution in all the simulations as the sources Lyman-$\alpha$ are near identical. After z$\approx$14 the Lyman-$\alpha$ becomes saturated. 

Our global heating histories are very different to those found in e.g. \citet{Semelin2017}. In their models significant ionization, heating and Lyman-$\alpha$ coupling occur considerably later than here. They find that typically this occurs below $z\sim10$, with peak around $z\sim8$. The main reason for these differences is that those simulations only include fairly massive sources, $M>10^{10}M_\odot$, which are very rare during reionization and only appear in significant numbers late.

\subsection{Evolution of the signal}

In Fig.~\ref{fig:lightcones:dbt}, we show the $\delta T_\mathrm{b}$ lightcones, slices from the position-redshift image cube, for each of the five simulations.  We show only late Lyman-$\alpha$ saturation, meaning that all models start with $\delta T_{\rm b}=0$. When compared to the results in Paper I, we see that the inclusion of the Lyman-$\alpha$ coupling effect in this case yields a weaker absorption signal, since heating is already underway before Lyman-$\alpha$ saturation ($y_\alpha \gg 1$) is reached. The inhomogeneous Lyman-$\alpha$ background also softens the features present before Lyman-$\alpha$ saturation ($z\approx17$). Unless stated otherwise, the late Lyman-$\alpha$ saturation case is the default.

\begin{figure*} 
\centering
\makebox[\textwidth][c]{\includegraphics[width=1.0\textwidth]{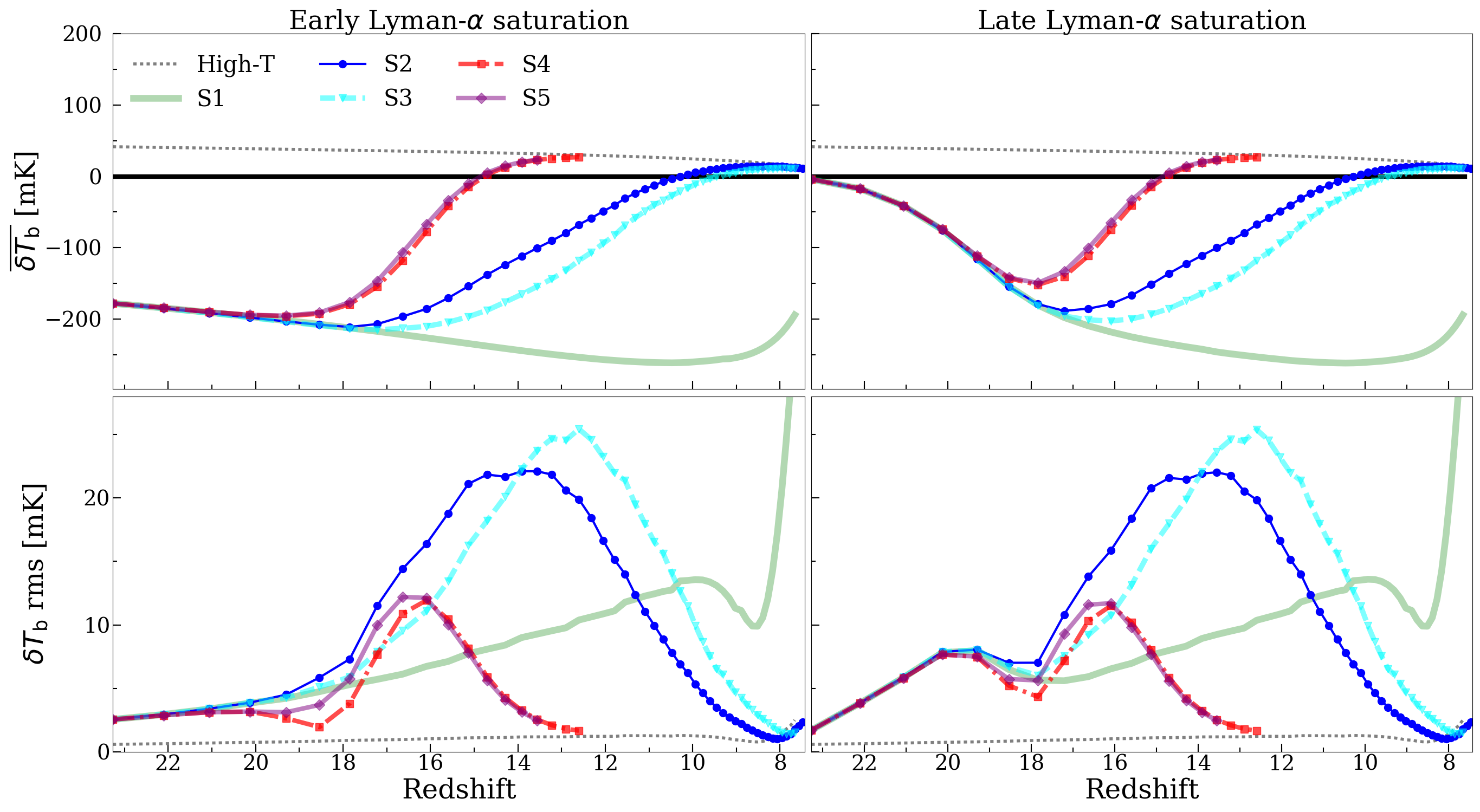}}
\caption{The mean and rms of $\delta T_{\rm b}$ for both early Lyman-$\alpha$ saturation (left-hand panels) and late Lyman-$\alpha$ saturation (right-hand panels). The high-$T_{\rm S}$ limit is shown to illustrate when temperature saturation occurs in each model.}
\label{fig:meanrms}
\end{figure*}

\begin{table} 
\begin{center}
\caption {\label{tab:noise} Table showing the expected noise on SKA1-Low from \citep{Koopmans2015} for a maximum baseline of 1.2~km and an integration time of 1000~h.}
\begin{tabular}{ | l | c | c | r |}
\hline
\ \ \ \ \ \ \ \ \ \ $z$ \ \ \ \ \  & \ \ \ \ 8.95 \ \ \ \ & \ \ \ \ 15.98 \ \ \ \ & \ \ \ \ 25.25 \ \ \ \   \ \ \ \ \ \\ \hline
\ \ \ \ \ \ \ \ \ \ $\theta$ [arcmin] \ \ \ \ \ & \ \ \ \ 6.0 \ \ \ \ & \ \ \ \ 10.3 \ \ \ \ & \ \ \ \ 15.8 \ \ \ \  \ \ \ \ \ \\ \hline
\ \ \ \ \ \ \ \ \ \ $\delta T_{\rm b}$ [mK] \ \ \ \ \ & \ \ \ \ 4 \ \ \ \ & \ \ \ \ 5 \ \ \ \ & \ \ \ \ 20 \ \ \ \ \ \ \ \ \  \\ 
\hline
\end{tabular}
\end{center}
\end{table}

After the Lyman-$\alpha$ background saturates, $\delta T_\mathrm{b}$ in simulation S1 remains in absorption for the rest of the simulation due to the lack of photons with long enough mean free paths to penetrate and heat the neutral IGM. Hence, the transition from absorption to emission never occurs in this case. At later times, ionized patches indicating the beginning of significant ionization can be seen as holes ($\delta T_\mathrm{b}\approx 0$) in the HI distribution.

In S2 and S3, the signal remains largely in absorption for longer than in S4 and S5, due to the total X-ray photon budget from the QSO sources being much lower than that of the HMXBs. However, at $z\approx16.5$, patches of higher $\delta T_{\rm b}$ start to develop around the QSO sources as they locally heat their surroundings. The sizes of these heated regions depend strongly on the spectra of the QSOs, with the harder spectrum S2 yielding noticeably more widespread heating. These regions are initially rare, but as more QSO sources form, they begin to overlap and bring the higher density regions into emission. Eventually, the QSOs become sufficiently numerous to heat the voids, and the entire simulation volume transitions into emission at $z\sim9.5$ and 9 for S2 and S3, respectively. In S2 and S3, early ionized bubbles are less sharply outlined, since the $\delta T_{\rm b}$ is closer to zero than it is in the other simulations. 

The X-ray heating in S2 and S3 is larger than that predicted in some earlier works. For example, \citet{Eide2018} find a negligible contribution to long-range heating from their QSO source model (referred in their paper as BH). Contrarily, the predictions from \citet{Yajima2014} and \citet{Datta2016} are more in agreement with our findings, both finding a more significant amount of X-ray heating for their single QSO.

In S4 and S5, the long-range X-ray heating due to HMXBs produces an earlier and less extended transition from absorption into emission at $z\sim14.5$ when compared to the QSO-only models (S2 and S3). Before temperature saturation is reached ($z\approx13.5$), large-scale  fluctuations throughout both S4 and S5 can be seen. These initial emission regions expand quickly, some increasing to tens of Mpc in size by $z\sim15.5$. Additional heated regions are visible in S5 at $z \approx 16$ where QSOs have formed; however, the transition from absorption to emission and temperature saturation happens at roughly the same time in these two models, as the heating from HMXBs dominates over that from the rare QSOs. These two cases follow a similar evolution to the results found in \citet{Mesinger2013} with X-ray efficiency $f_x=1$, although significant reionization begins earlier in their model. On the other hand \citet{Semelin2017} find that, in all models, their signals transition to emission much later and is much more patchy than our model due to their rarer sources.

In Fig.~\ref{fig:blurred}, smoothed mean-subtracted maps of the 21-cm signal are displayed. To generate these maps coeval cubes are smoothed in the two angular directions with a two-dimensional Gaussian beam with a FWHM corresponding to a 1.2~km baseline at the frequency corresponding to the redshift of interest ($\theta_\mathrm{FWHM}=0.221(1+z)/1200$). This maximum baseline length approximately corresponds to the planned size of the core of SKA1-Low. Along the line-of-sight (frequency) direction, the data is smoothed with a top-hat function, the width of which is equal to the co-moving distance corresponding to $\theta_\mathrm{FWHM}$.  Table~\ref{tab:noise} lists the angular resolution together with the expected rms noise value at this resolution for three representative redshifts.

At $z\sim16$, the SKA1-Low noise levels for deep integrations and our standard resolution are expected to be around 5~mK. From the first column of Fig.~ref{fig:blurred} we can see that all our simulations show fluctuations that exceed this value. S4 and S5 exhibit peak fluctuations around 20~mK, and S2 and S3 have even higher ones, reaching over 40~mK in magnitude, at lower redshifts. These levels imply that each of these X-ray source models could possibly be imaged directly with SKA1-Low.  Models including HMXBs, S4 and S5, then approach temperature saturation and become indistinguishable from each other after $z\sim15$. Before this, heating from QSOs could be seen in S5, in agreement with \citet{Yajima2014} and \citet{Datta2016}, which both conclude that their single QSO source may be detectable by SKA due to the large emission region it produces.

Fig.~\ref{fig:blurred} shows that all our models can be distinguished visually at $z\sim16$, showing the sensitivity of the 21-cm signal from the CD to the presence different sources. However, it should be noted that we have only explored a very small part of the enormous parameter space and there may well be degeneracy between the source parameters and the 21-cm signal they produce. For example, different combinations of emissivities and X-ray spectra for the same source types could feasibly produce indistinguishable 21-cm maps. 

In Fig.~\ref{fig:meanrms}, we show the evolution of the mean $\delta T_\mathrm{b}$ (top panels) and its rms fluctuations, smoothed to the expected SKA resolution (bottom panels) for both early (left panels) and late (right panels) Lyman-$\alpha$ saturation. In all cases, we also include the high-$T_\mathrm{S}$ limit case, indicating when the temperature saturation limit is being approached.

When we assume Lyman-$\alpha$ saturation (top left panel), all simulations start in absorption due to the initially very cold IGM. In S1, the IGM remains cold and thus $\overline{ \delta T_\mathrm{b}}$ decreases all the way to $z\sim11$ when the highest density peaks start to become ionized and cause the signal to increase slightly.

The signal in S2 and S3 initially follows a similar pattern; however, after redshift $z\approx17$, heating from the QSOs begins to have an impact and gradually raises the temperature of the neutral IGM until it eventually asymptotes to the high-temperature limit at $z\approx8-9$. The more energetic QSOs in S2 contribute more to heating and produce a weaker absorption signal and earlier transition to emission than those in S3. The presence of HMXBs has a much greater impact on the global mean than the QSOs for the models chosen here. Consequently, the mean value from S4 begins to increase markedly earlier and follows a patern more similar to that seen in \citet{Mesinger2013}. Our models predict that the value of $\overline{\delta T_{\rm b}}$ rises significantly earlier than all models shown in \citet{Semelin2017}, again due to the relative rarity of their X-ray sources.

In the case of late Lyman-$\alpha$ saturation (top right panel), $\overline{ \delta T_\mathrm{b}}$ instead starts at zero, as the Lyman-$\alpha$ coupling is inefficient at these early times. This mean value then decreases to meet the fully coupled cases, as the Lyman-$\alpha$ background appreciates. Late Lyman-$\alpha$ coupling considerably reduces the length of the period of strong absorption compared to the case of early Lyman-$\alpha$ saturation. The absorption signal peak magnitude is noticeably reduced in the early-heating cases S4 and S5. In contrast, the magnitude is is unaffected in the late-heating cases, as Lyman-$\alpha$ saturation is achieved before the absorption signal peaks. \citet{Semelin2017} find that their rarer sources produce a Lyman-$\alpha$ background that saturates much more gradually.

The lower panels of Fig.~\ref{fig:meanrms} show the rms (i.e the standard deviation) of $\delta T_\mathrm{b}$, smoothed to the expected resolution of SKA. The lower left-hand panel of Fig.~\ref{fig:meanrms} displays the rms calculated for the case of early Lyman-$\alpha$ saturation. Before the X-ray heating is able to have significant impact on the cold IGM, all scenarios follow a similar rms evolution, which is dominated by the density fluctuations and the adiabatic cooling of the IGM. The fluctuations in S1 continue to be driven by density fluctuations, which increase as structure formation progresses. At $z\sim10$, ionized bubbles begin to grow around the sources, increasing $\delta T_{\rm b}$ in dense regions near sources. As the deep absorption signals from these dense regions become weaker the rms decreases. As these bubbles continue to expand, $\delta T_{\rm b}$ approaches zero in increasingly large regions around the sources, contrasting with the signal from the neutral IGM and increasing the rms. The high-$T_{\rm S}$ limit follows the same evolution, but the rms fluctuations are much lower due to the signal being assumed to be in emission rather than absorption.

The rareness of QSOs in S2 and S3 introduces large-scale heating fluctuations, which increase the peak rms values by a factor of about two compared to S4 and S5, before the rms decreases in all heating scenarios as heating saturation is approached. The softer QSO model S3 peaks somewhat later than S2 as the harder QSOs produce more energy in our models. The additional heating fluctuations due to the QSOs yield a slightly earlier rms fluctuations peak in S5 ($z \approx 16.5$) compared to S4 ($z \approx 16$).

In all simulations, the inclusion of the inhomogeneous Lyman-$\alpha$ (Fig.~\ref{fig:meanrms}, bottom right) background boosts the early fluctuations, which results in an additional peak in the rms at $z\sim20$. These additional fluctuations come from inhomogeneities in the Lyman-$\alpha$ background being introduced into the signal. Fluctuations at later times ($z<18$) are unaffected, since the Lyman-$\alpha$ background saturates. 
\begin{figure*} 
\centering
\includegraphics[width=1.0\textwidth]{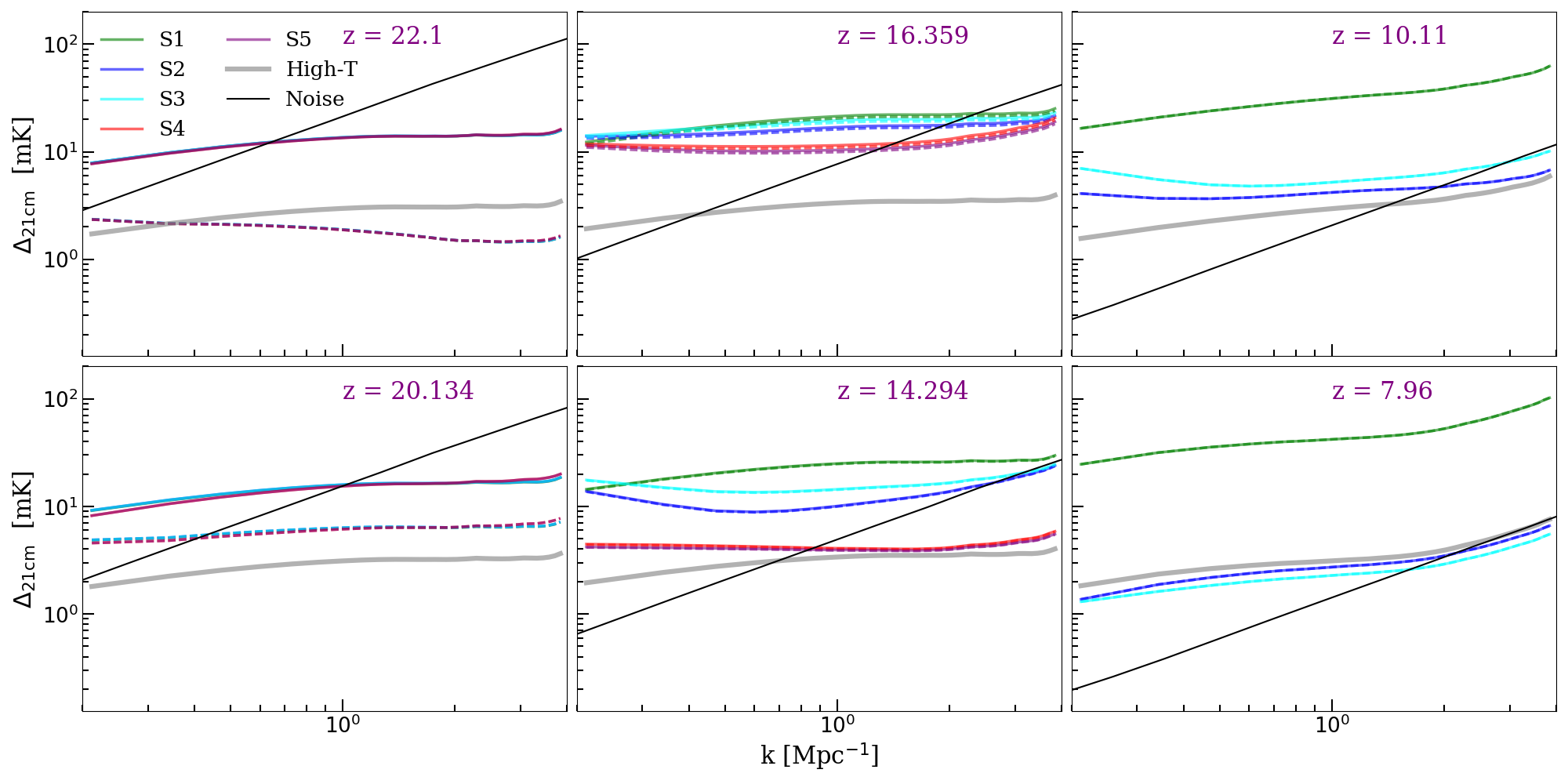}
\caption{The 21-cm power spectra from our simulations at several key stages of the evolution assuming early (solid lines) and late (dashed lines) Lyman-$\alpha$ saturation. For reference, the high-$T_\mathrm{S}$ limit is indicated, as labeled. The error due to noise is also included (straight lines, as labelled).}
\label{fig:powerspectra}
\end{figure*}

\subsection{Power Spectra}

In Fig.~\ref{fig:powerspectra}, we show the power spectra of $\delta T_\mathrm{b}$ at several key redshifts along with the error on the power spectra due to noise (calculated as outlined in \citet{Villaescusa-Navarro2014}). At the beginning of the simulation ($z\sim22$), all simulations give identical results with very high (absorption) signal for early Lyman-$\alpha$ saturation and very low signal for late Lyman-$\alpha$ coupling, since in the latter case, the 21-cm signal is still coupled to $T_{\rm CMB}$. At this stage, the heating has not yet had a significant impact. Thus in the early Lyman-$\alpha$ saturation cases, the 21-cm fluctuations simply follow the density ones, with a boost due to the strong absorption compared to the high-$T_{\rm S}$ limit.  

If the Lyman-$\alpha$ background has not yet saturated, the power is suppressed on all scales. As the Lyman-$\alpha$ couples to the CMB in the regions close to the stars first, the large distances between these regions boost the power on large scales. However, this conclusion appears dependent on the Lyman-$\alpha$ model employed. Early work \citep{Santos2008} found the opposite effect. However, model S7 in \citet{Baek2010} yields a power spectra with a similar shape to ours, albeit with more power especially on smaller scales. As they have plotted power spectra as a funciton of $x_\alpha$ rather than redshift this difference in magnitude could be due to comparing different redshifts.

By $z\sim20$, Lyman-$\alpha$ coupling has become efficient in the late Lyman-$\alpha$ saturation case. The power has increased on all scales compared to the beginning of the simulation, particularly the very large scales. There is more power at large scales in the late Lyman-$\alpha$ saturation case than the early Lyman-$\alpha$ saturation case due to the large-scale fluctuations introduced by the inhomogeneous Lyman-$\alpha$ background. 

\citet{Santos2008} find the inverse to this, with fluctuations from Lyman-$\alpha$ being introduced on small rather than large scales at this redshift. \citet{Pritchard2007} find that Lyman-$\alpha$ fluctuations dominate at this stage, due to the different timing of their models. In the case of late Lyman-$\alpha$ saturation, the models remain almost the same, with only a slight difference between the models with and without HMXBs. \citet{Santos2008} find comparable results to our own. Due to the different timings of the models, it is difficult to compare to the results of \citet{Semelin2017}. We compare this redshift to their power spectra from z$\approx$10 (where their heating has just begun and Lyman-$\alpha$ is not yet fully coupled). These power spectra agree on large scales, but we find somewhat more power on smaller scales. This is likely due to the presence of more less massive sources in our simulations, as they introduce fluctuations on smaller scales.

\begin{figure*} 
\includegraphics[width=1.\textwidth]{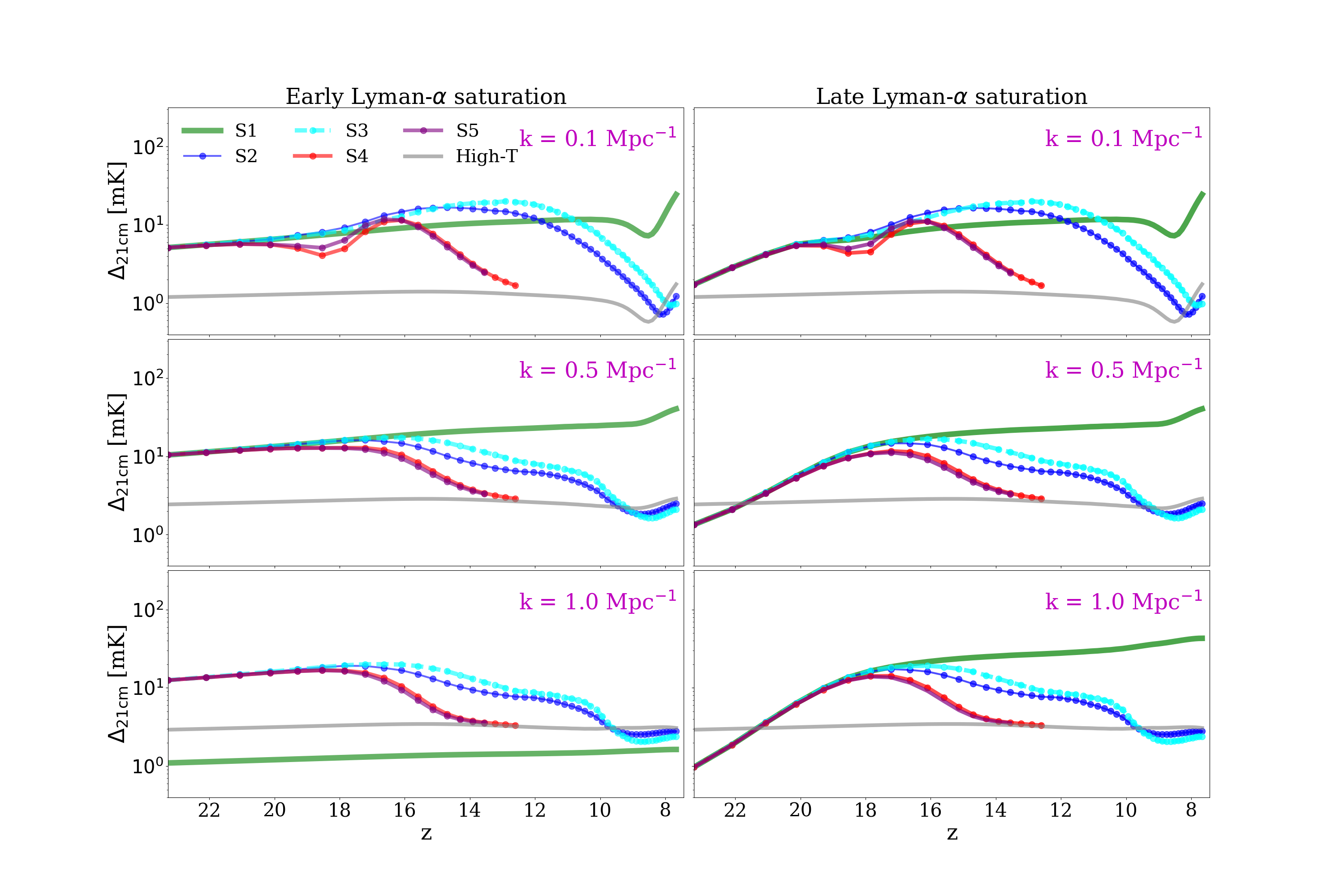}
\caption{The evolution of the 21-cm power spectra modes at $k=0.1,0.5$, and $1\,$Mpc$^{-1}$ for all X-ray source models. On the left-hand side panels, power spectra modes from the early Lyman-$\alpha$ saturation scenario are displayed, and on the right, the late Lyman-$\alpha$ saturation. The high-$T_{\rm S}$ limit is displayed to illustrate temperature saturation.\label{fig:kplot}}
\end{figure*}

By $z\approx 16$, the Lyman-$\alpha$ background is sufficiently built up so that the early and late Lyman-$\alpha$ saturation results converge. This is somewhat earlier than found in \citet{Santos2008}, where Lyman-$\alpha$ saturation does not occur until z$\approx$10. The small-scale power in simulations S4 and S5 begins to decrease, as X-ray heating washes out small-scale temperature fluctuations; whereas on larger scales, the power is slightly boosted for the same reason. S2 and S3 follow a similar pattern, but have more power on large scales due to the rareness of the brighter QSOs. On smaller scales, they also have more power, as significant heating has not extended to much of the simulation and hence has not washed out the small-scale temperature fluctuations from previous QSO activity. Both \citet{Santos2008} and \citet{Baek2010} find power spectra with a similar magnitude and shape at this time, but with a peak at smaller scales than those found here.

\begin{figure*} 
\centering
\makebox[\textwidth][c]{\includegraphics[width=1.0\textwidth]{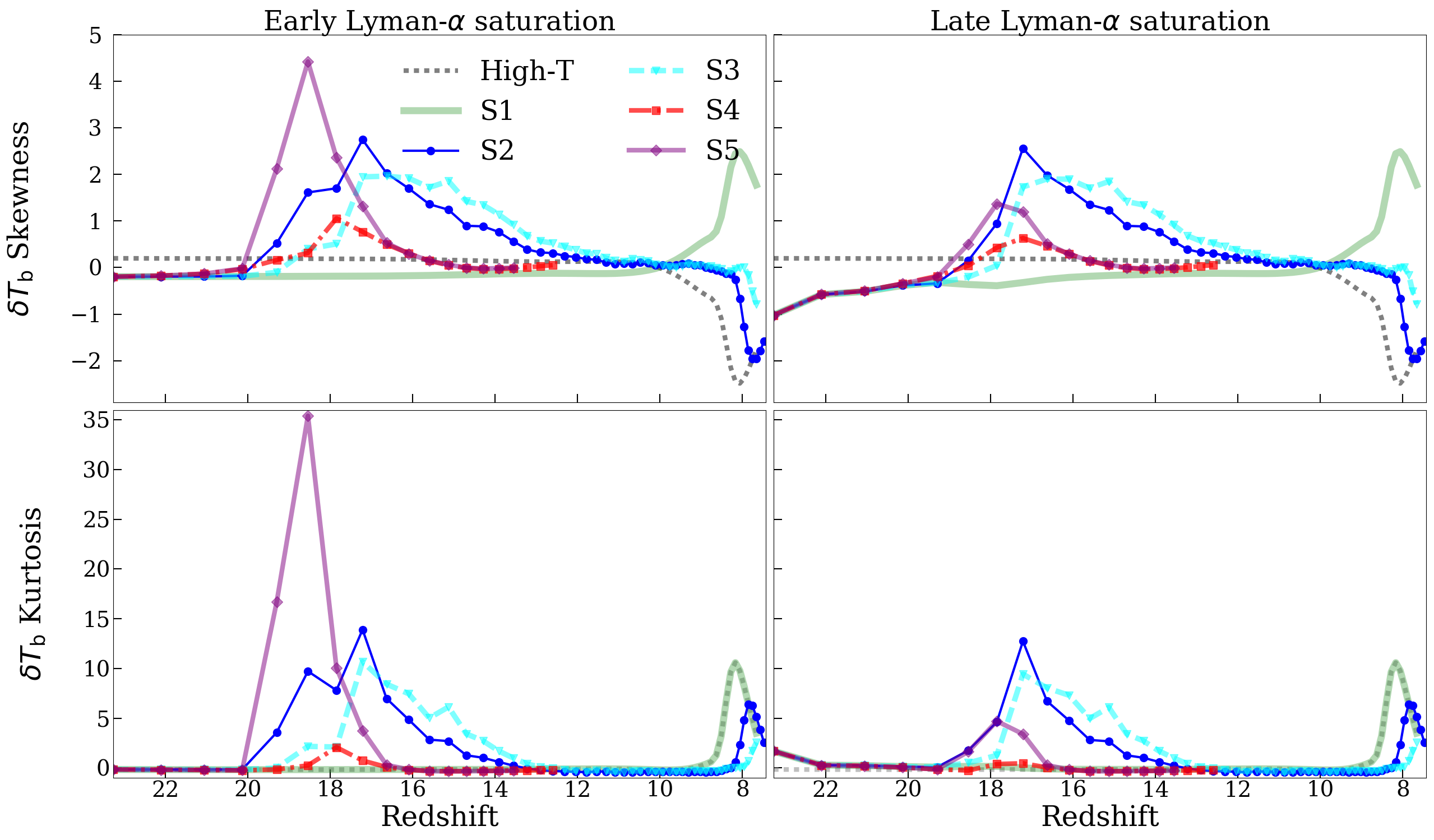}}
\caption{The evolution of the skewness (top panels) and kurtosis (bottom panels) of $\delta T_{\rm b}$ for all simulations, as labeled, for early Lyman-$\alpha$ saturation (left panels) and late one (right panels).}
\label{fig:statistics}
\end{figure*}

By $z\approx 14$, the 21-cm power is near its peak in S2 and S3 due to large-scale heating fluctuations, introducing strong contrast between the hot and cold regions. The power is starting to decline significantly in S4 and S5. As the transition to emission takes place, convergence to the high-$T_{\rm S}$ limit (which can then be used to describe the rest of the EoR accurately) is approached. The power in S2 and S3 is still boosted on large scales compared to that of S1, but long-range heating is beginning to suppress the power on smaller scales. 

At $z\approx 10$, S4 and S5 have reached temperature saturation. These models are in rough agreement with the power spectara found in \citet{Santos2008}. S3 and S4 approach temperature saturation much more gradually and later ($z \approx$ 7.9), with slightly more power remaining in large-scale fluctuations. S3 evolves more slowly than S2 due to the QSOs producing less energy in this model, thus heating more locally. Power on all scales in S2 and S3 is lower than in the high-$T_{\rm S}$ case, as reionization has begun. Regions of the simulation that are transitioning from absorption to emission have values closer to the zero-signal coming from ionized regions, which decreases the magnitude of the fluctuations. In the fully saturated case, there is more contrast between the emission signal from heated, neutral regions and zero signal from ionized regions, so the magnitude of fluctuations is greater.

The redshift evolution of several $k$-modes are shown in Fig.~\ref{fig:kplot}. Panels on the left-hand side show results from the early Lyman-$\alpha$ heating. On large scales, X-ray heating from HXMBs initially suppresses fluctuations, as heating weakens the absorption signal from the densest regions around sources. As these regions move towards emission, the large-scale fluctuations are boosted before decreasing once more, as temperature saturation approaches. When in combination with HMXBs, QSOs follow a similar pattern. However, with QSOs alone, there is no trough present as many high density regions remain in strong absorption due to the rarity of the QSOs. 

The evolution on smaller scales is somewhat similar, however, at no point are the fluctuations boosted by X-ray heating. The removal of deep absorption signals from dense regions around sources causes this decrease in power, occurring more rapidly in models with HMXBs as X-rays are emitted from each halo. Power on small scales is suppressed more rapidly in S2, as long-range heating is more significant.

Fluctuations including the inhomogeneous Lyman-$\alpha$ background from late Lyman-$\alpha$ saturation are shown in the right-hand panels. Initially, fluctuations are suppressed in all models until the Lyman-$\alpha$ background has been established. This Lyman-$\alpha$ case particularly impacts the fluctuations in S4 and S5, decreasing the peak value at $z=16$ to log$_{10}(\Delta_{\rm 21cm})\approx 0.9$. The peak values of S2 and S3 occur later, after Lyman-$\alpha$ saturation is reached and so are unaffected.

\begin{figure*} 
\centering
\makebox[\textwidth][c]{\includegraphics[width=1.0\textwidth]{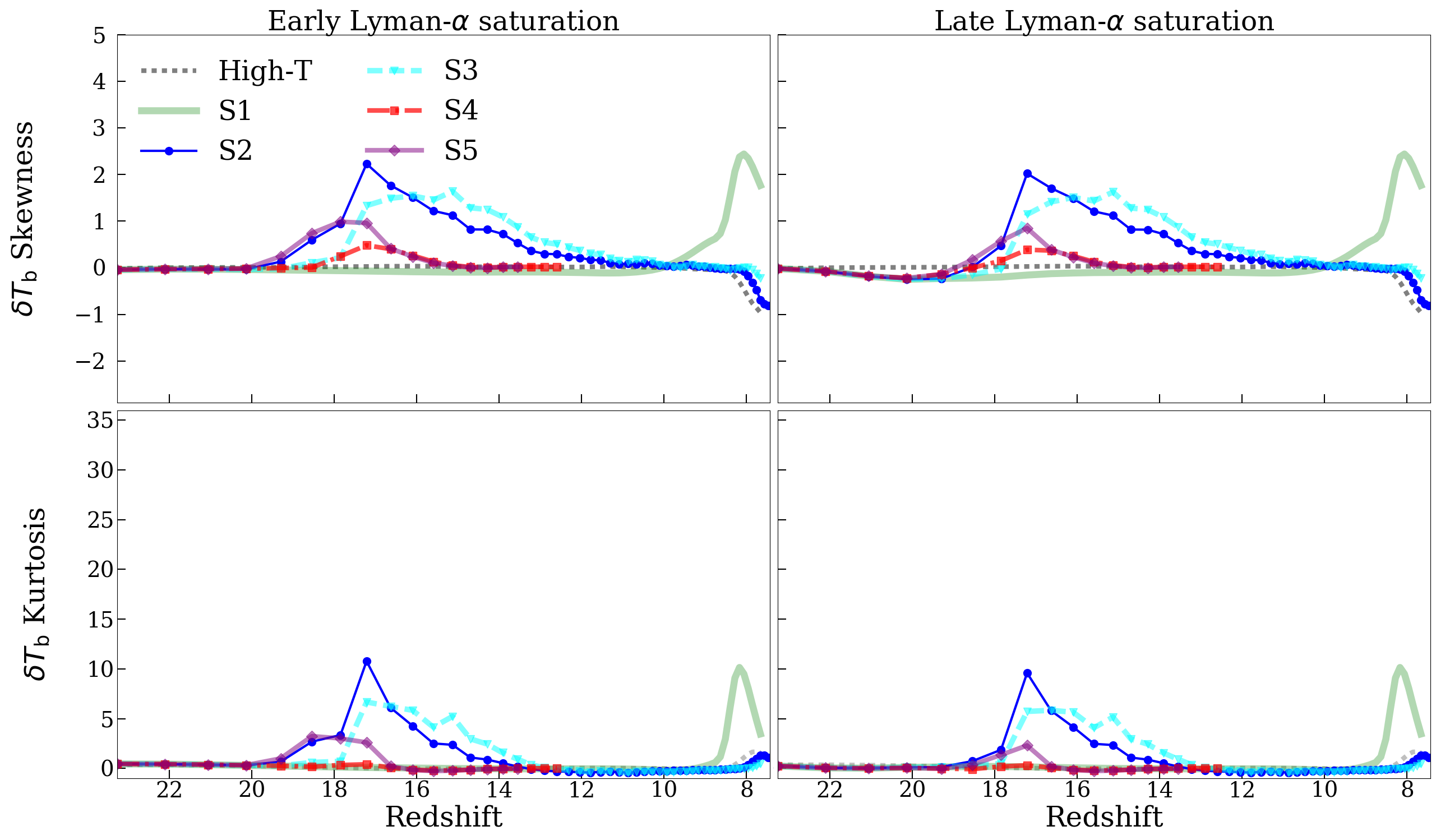}}
\caption{The evolution of the skewness (top panels) and kurtosis (bottom panels) of $\delta T_{\rm b}$ with telescope noise added. Results from all simulations are shown for early Lyman-$\alpha$ saturation (left panels) and late one (right panels).}
\label{fig:statisticsnoise}
\end{figure*}

\subsection{Non-Gaussianity of the 21-cm signal}

The power spectra alone cannot be used to fully describe $\delta T_{\rm b}$ fluctuations from X-ray heating as they are highly non-Gaussian. Therefore, we consider the higher moments (skewness and kurtosis) of the one-point statistics of the 21-cm signal produced from our simulations. We use the following dimensionless definitions for skewness and kurtosis:
\begin{equation}
\mathrm{Skewness}(y) = \frac{1}{N} \frac{\sum_{i=0}^{N}(y_i - \overline{y})^3}{\sigma^{3}},
\label{eq:skew}
\end{equation}
and
\begin{equation}
\mathrm{Kurtosis}(y) = \frac{1}{N} \frac{\sum_{i=0}^{N}(y_i - \overline{y})^4}{\sigma^4}.
\label{eq:kur}
\end{equation}
Here, $y$ is the quantity of interest (i.e., $\delta T_{\rm b}$), $N$ is the total number of data points, and $\overline{y}$ and $\sigma^2$ are the mean and variance of $y$, respectively. These quantities are all smoothed to the resolution of SKA1-Low and are calculated from coeval simulation boxes. The fact that QSO sources are stochastic and do not trace the Gaussian density distribution leads to a dramatic increase in non-Gaussianity. Therefore, compared to the other measures discussed, these higher order statistics demonstrate the most extreme differences between models with and without QSOs.

In Fig.~\ref{fig:statistics}, we show the evolution of the skewness (top) and kurtosis (bottom) for both early (left) and late (right) Lyman-$\alpha$ saturation. S1 shows a flat, featureless evolution throughout the CD, since the density field at early times is close to Gaussian. The skewness only increases once significant reionization begins at $z\sim10$, when stars ionize the high-density peaks. These ionized regions introduce non-Gaussianity into the signal, as shown in previous studies \citep{Mellema2006}. 

In all cases with X-ray heating, the skewness initially follows that of S1 until $z\sim20$ when QSOs begin to form in S2, S3, and S5. In the hard-spectra QSO cases S2 and S5, the skewness increases rapidly; while in the softer-spectrum case S3, it increases more gradually. Similarly in S4, the skewness increases gradually (and peaks at a lower value), since non-Gaussianities are added by the (softer-spectrum) HMXBs. The maximal skewness from S5 is 4.5, or over four times greater than the value ($\sim1$) obtained from S4. The peak in S5 also occurs somewhat earlier, at $z\sim18.5$ rather than $z\sim18$. The maximal skewness values obtained from S2 (2.7) and S3 (2.0) are intermediate between those from S4 and S5, while still much higher (by factors of 27 and 20, respectively) than the value found for S1 (0.1). The largest value for S2 occurs earlier (at $z\sim17.5$) than that of S3 ($z\sim16$), as the harder QSOs heat more rapidly the cold IGM patches responsible for driving the 21-cm signal fluctuations. 

The inhomogeneous Lyman-$\alpha$ background (Fig.~\ref{fig:statistics}, right panels) introduces additional non-Gaussian fluctuations of $\delta T_{\rm b}$ due to regions around the sources that first become decoupled from $T_{\rm CMB}$. These dominate over the density fluctuations and give rise to the negative skewness observed in all cases, which then gradually increases as the Lyman-$\alpha$ background builds up as more of the simulation volume decouples from $T_{\rm CMB}$. Thereafter ($z<19$), heating starts, and temperature variations begin to impact the signal. In all cases, the skewness increases and eventually peaks; however in the X-ray source models where heating occurs earlier, the peaks due to heating fluctuations are suppressed compared to the early Lyman-$\alpha$ saturation. This suppression is particularly evident for S5.

The kurtosis (Fig.~\ref{fig:statistics}, lower panels) follows the same qualitative pattern, but with an even more extreme difference between the QSO and non-QSO cases. The maximal value of the kurtosis from S4 ($\sim1$) is 17 times smaller than the corresponding value from S5. The kurtosis yields a more notable difference between the two QSO models, with S2 reaching a peak value of 14 versus 10 for S3 (both are significantly greater than the value from S1, which is close to zero). The greater amount of heating from the harder QSOs in S2 again leads to a larger deviation from Gaussian fluctuations observed.

When Lyman-$\alpha$ coupling is not complete before heating begins (Fig.~\ref{fig:statistics}, bottom right), the maximal kurtosis is also suppressed. This effect is most notable for the X-ray source models S4 and S5, where heating occurs earlier, with the peak in S4 being totally suppressed. The maximum value of S5 is decreased by a factor of 10. The kurtosis of S2 and S3 are slightly suppressed initially, but the largest values remain the same as they occur at lower redshift when Lyman-$\alpha$ is closer to saturation. 

Finally, in Fig.~\ref{fig:statisticsnoise} we show the skewness and kurtosis of our simulations with the addition of telescope noise using the method outlined in \citet{Ghara2017} and \citet{Giri2018}. The strong non-Gaussianity in the signal is somewhat washed out, especially at out at higher redshifts. The peaks values of both the skewness and the kurtosis are lowered, particularly in the case of early Lyman-$\alpha$ saturation. The non-Gaussianity from reionization itself is also decreased. However, even when telescope noise is present our different Cosmic Dawn X-ray models can still be distinguished from each other.

\section{CONCLUSIONS}
\label{sec:conclusions}

In this paper, we present a suite of large-volume, fully numerical radiative transfer simulations of X-ray heating of the IGM during the CD, extending our previous work in Paper I. We introduced two types of QSO sources, with power-law spectral slopes of -0.8 and -1.6, and compare their impact to the effects of HMXB X-ray sources considered in Paper I, as well as a new case combining QSO and HMXB sources and a fiducial, stars-only simulation. Unlike HMXBs, the QSOs are rare and are assigned randomly to HMACH haloes with luminosities sampled from the high-redshift extrapolation of an empirical QXLF. These luminosities are not proportional to the host halo mass. As the precise nature and properties of these early sources remain uncertain, we have chosen a QXLF that predicts fairly numerous QSOs in order to examine their maximum possible impact. Our simulations show QSO sources may be able to contribute non-trivially to early X-ray heating and also suggest that it is possible to distinguish between soft- and hard-spectra models using the resulting 21-cm signal, particularly via the non-Gaussianity of the signal. 

These QSO sources contribute many fewer photons to the X-ray heating than HMXBs, so their overall energy contribution is subdominant compared to HMXB sources when both source types are present. On their own, both QSOs models yield a considerably more extended transition of the 21-cm signal from absorption to emission, and spin temperature saturation of the neutral IGM is not reached until reionization itself is well under way. The late temperature saturation in the simulations with only QSOs cause the non-Gaussianity from reionization itself ($z<10$) to be lower than in the case of full saturation of the spin temperature. This effect is more pronounced for the QSOs with softer X-ray spectra.

During the CD, heating from QSOs has a more notable impact on the heating fluctuations than on the mean value of $\delta \rm T_b$. When compared to the stellar-only case, the $\delta \rm T_b$ power spectrum from all X-ray models show more power on larger scales and less on smaller scales until around $z\sim16$. After this time, the power spectrum of simulations including HMXBs decreases on all scales as temperature saturation is approached, but the power spectra for the QSO cases experience a further boost on large scales.  The rms fluctuations for all X-ray source models are above the expected noise levels for observations with the SKA1-Low core, implying that low resolution tomographic imaging of the CD may be possible. The rare QSOs boost the rms, particularly when they are the only sources of X-ray heating. In this case, the peak value of the fluctuations are $\sim$10~mK higher than in the HMXB cases, and this peak occurs at lower redshift.

By far, the clearest signature of the QSOs is found in one-point higher order statistics of the 21-cm signal PDF distribution: skewness and kurtosis. An increase in both quantities can be seen both when QSOs are the sole sources driving X-ray heating and when they are present with HMXBs. These strong non-Gaussianities are driven by the rareness of the QSOs, introducing fluctuations in the signal largely unrelated to the underlying (mostly Gaussian) density field. However, this increase in non-Gaussianity can be partly suppressed by late Lyman-$\alpha$ saturation, so while an extremely non-Gaussian signal from the CD could indicate the presence of QSOs, a more Gaussian signal would not rule them out. 

In addition to suppressing the non-Gaussianities, the Lyman-$\alpha$ background fluctuations (in all models) produced by late Lyman-$\alpha$ saturation cause an additional peak in $\delta \rm T_b$ fluctuations, as has been found in several previous works \citep[e.g.][]{Santos2008,Baek2010,Ghara2015,Watkinson2015}. The power spectra show that this is due mainly to contribution from larger scales. The peak magnitude of these rms fluctuations driven by the Lyman-$\alpha$ background is $\approx$8~mK, which is well below the expected image noise of 20~mK for imaging with the SKA1-Low core. However, a power spectrum detection of these fluctuations should still be possible.

The difference found between our work and that in \citet{Eide2018} is model dependent and is in part due to their seeding algorithm only allowing the formation of black holes in haloes greater than 10$^{10}$M$_\odot$, resulting in a much lower number density of QSOs. This lower density combined with the assumption that QSOs have optically thick, geometrically thin disks (as described in \citet{Shakura1973}) leads to a somewhat more conservative heating prediction than in our models. This is illustrated by our agreement \citet{Yajima2014}, who follow a prescription similar to \citet{Eide2018} but assumes that black holes accrete at their Eddington luminosities. \citet{Datta2016} also predict that their individual, bright QSO would be detectable in 1000~h integrations with SKA1-Low -- which is in agreement with our own predictions. \citet{Datta2016} use observations of the low-redshift QSOs to determine their luminosities, a method more comparable to our own.

A key implication of the results presented in this work is that all the X-ray source models we investigated go through phases in which the fluctuations in the 21-cm signal are above the expected noise for observations with the core of SKA1-Low \citep{Koopmans2015}, suggesting that at least part of this epoch could not only be studied with power spectra, but directly imaged. In Paper~I, we found this to be the case for HMXBs, and we can now extend this conclusion to the cases with rare, QSO-like sources. With higher levels of galactic foregrounds and stronger ionospheric effects, imaging around $z\sim 16$ will not be easy, but images of the CD would open the door to a multitude of analysis techniques to extract information from the signal about the astrophysics of the CD. Examples include parameter estimation through deep learning of images \citep[e.g.]{Shimabukuro2017,Gillet2018} and MCMC approaches \citep[e.g.]{Greig2018}, emulators \citep[e.g.]{Kern2017}, the bispectrum \citep[e.g.]{Shimabukuro2016}, and size distributions of features \citep[e.g.][]{Giri2018}. 

\citet{Ghara2015} and \citet{Baek2010} find power spectra with similar magnitudes to our own. Detailed comparisons to these works are complicated by the fact that \citet{Baek2010} has no subgrid model for unresolved, low-mass sources and \citet{Ghara2015} use a subgrid model very different from the one employed in this work. The subgrid modeling is particularly important for comparisons as lower resolutions of these simulations mean that resolved haloes do not appear until later in the CD. \citet{Pritchard2007} and \citet{Pacucci2014} also find power spectra with comparable magnitudes. The large-scale fluctuations in these works (at $k\approx 0.1$Mpc$^{-1}$) peak at roughly the same time as our models including HMXBs.

\citet{Baek2010} and \citet{Watkinson2015} include measures of the power spectra and higher order statistics of the 21-cm signal produced by HMXB sources. The peak skewness values during the CD found from models S4 and S2 in \citet{Baek2010} are in agreement with the ones found in our HMXB model (S4), but are significantly lower than the values found from our simulations including QSOs. The lack of correlation between our QSO luminosities and their host dark matter haloes, along with their rareness are the factors driving this. The difference is far more pronounced in the early Lyman-$\alpha$ saturation case, but still noticeable in the late Lyman-$\alpha$ saturation case. Similarly, the skewness found for all models in \citet{Watkinson2015} have peak values far lower than that of our QSO models (but the case `$\log\zeta_{\rm X}=55$' is in agreement with our HMXB model). 

The high level of non-Gaussianity produced by QSOs imply that higher order statistics may be a more useful probe than the power spectra to discriminate between certain source models in the CD. In particular, high non-Gaussianity likely indicates the presence of rare sources, such as QSOs. The skewness shows a clear difference between our models including QSOs and our models containing HMXBs, as well as models from other works. Clearly our QSO-like sources introduce a significantly greater amount of non-Gaussianity in the signal. This result further motivates the use of alternative analysis techniques to interpret the signal from the CD to probe for rare X-ray sources. The bispectra of the 21-cm signals from all the X-ray simulations presented here have been extensively studied in \citet{Watkinson2018}.

Conversely, investigating the non-Gaussianities during the EoR could be less insightful than previously hoped if late heating occurs. Despite the additional fluctuations introduced to the kinetic temperature of the gas, non-Gaussianties during the EoR are in fact lower in our QSO-only models. This decrease is due to the magnitude of $\delta T_{\rm b}$ from the neutral regions being lower in the late heating case than for the saturated spin temperature case, removing the brightest points from the signal. 

There are some other potential X-ray sources that may have contributed to the fluctuations in the 21-cm signal in the CD that we have not yet considered, for example supernovae \citep[e.g.][]{Yajima2015}. In addition, there is still a large parameter space associated with our current sources, for example varying the star formation efficiency of haloes and hence the luminosity of our HMXBs. Due to the computational expense of our simulations, we have not yet been able to fully explore the huge parameter space associated with the CD. However, the alternative of exploring the parameter space with fast semi-numerical models may not be sufficient. Ideally, these two approaches should be combined in order to achieve reliable predictions and interpretation of any observational detections.

Another limitation of the current work is the modeling of the Lyman-$\alpha$ background. We have included only the two most extreme cases, one where very early sources build up a Lyman-$\alpha$ background before the simulation begins (early Lyman-$\alpha$ saturation) and one where only sources present in our simulation volume contribute to the Lyman-$\alpha$ background (late Lyman-$\alpha$ saturation). However, the most likely scenario is somewhere in between, with earlier sources, such as mini-haloes, contributing a non-negligible Lyman-$\alpha$ flux to the background, but not achieving full Lyman-$\alpha$ saturation.

Finally, recent observations suggest that the z$\approx$10 Universe may not be as dust and metal free as previously thought \citep{Chiaki2019,Tamura2019}. Dust impacts the properties and evolution of galaxies, in particular the escape fraction, which could impact our results for this time. Our RT calculations assume hydrogen and helium to be the only elements present and may not be sufficient once enough metal enrichment has occurred. We leave these considerations for future work.

Despite these few caveats, the results from our simulations have shown that the 21-cm signal from the CD may not only be statistically detectable with SKA1-Low, but also imageable. Our X-ray source models have distinct power spectra with markedly different the evolution for our difference models. Finally, the high non-Gaussianity driven by X-ray heating illustrates the need to consider statistics beyond the power spectrum when considering this signal, particularly when considering rare X-ray sources.

\section{Acknowledgements}
This work was supported by the Science and Technology Facilities Council [grant numbers ST/I000976/1 and ST/P000525/1] and the Southeast Physics Network (SEPNet). GM is supported in part by Swedish Research Council grant 2016-03581. This research was supported in part by the Munich Institute for Astro- and Particle Physics (MIAPP) of the DFG cluster of excellence ``Origin and Structure of the Universe". We acknowledge that the results in this paper have been achieved using the PRACE Research Infrastructure resource Marenostrum based in the Barcelona Supercomputing Center, Spain under Tier-0 project 'Multi-scale Reionization'. We acknowledge that some of the results of this research have been achieved
using the DECI resource Cartesius based in Netherlands at SURFSara with support from the PRACE aisbl. Some of the numerical computations were done on the Apollo cluster at The University of Sussex. Part of the simulations were performed on resources provided by the Swedish National Infrastructure for Computing (SNIC) at the PDC Center for High Performance Computing in Stockholm. The $N$-body simulation used in this work was completed under the Partnership for Advanced Computing in Europe, PRACE, Tier-0 project PRACE4LOFAR on the TGCC Curie computer. Finally, we would like to thank the referee for helping us improve this work with their constructive comments.

\bibliography{paper}

\appendix
\section{Multiphase algorithm}
Many mesh based calculations adopt the finite-volume approach, which means that the value of a quantity $Q$ inside a cell is the average of this quantity over the volume of the cell $V$
\begin{equation}
  \langle Q \rangle = \int_V Q \mathrm{d}V\,.
\end{equation}
Problems arise when derived quantities rely non-linearly on one or more calculated quantities $Q_i$, as generally
\begin{equation}
  \langle Q^k \rangle \neq \langle Q \rangle^k\, 
\end{equation}
if $k\neq 1$. Examples of this in the context of photoionization calculations are the recombinations rates and collisional cooling rates (both are proportional to $n^2$, where $n$ is the density). In the context of recombination rates, this discrepancy can sometimes be corrected for by using clumping factors
\begin{equation}
  C=\frac{\langle n^2 \rangle} {\langle n \rangle^2}\,,
\end{equation}
if the density variations inside a cell are known.

The width of ionization fronts (I-fronts) is approximately 20 photon mean free paths, which for soft (low-energy) photons is typically much smaller than the spatial resolution in cosmological-volume simulations. Therefore, the I-front transition is quite sharp and cannot be easily resolved. When the radiative transfer code does not resolve ionization fronts, some cells will be partly inside an ionized region, where the hydrogen ionization fraction $x\approx 1$, and partly outside, where $x \approx 0$. Let us assume that a fraction $f$ of a cell is inside and $1-f$ outside the ionized region. Such a cell can be described as multiphase, as it contains both an ionized phase and a neutral phase. Since for pure hydrogen the recombination rate is proportional to $n(\mathrm{HII})n(e)$ and $n(\mathrm{HII})=n(\mathrm{e}^-)=xn$, the average recombination rate in the cell will be proportional to $fn^2$; whereas, the value derived from the finite volume values of the cell will be $f^2n^2$, as $\langle n(\mathrm{HII}) \rangle = fn$. This error is usually ignored as it is often transient, around the time the I-front passes through the cell, and relatively small. However, for large cells and weak sources, cells may be in a multiphase state for a long time, and the cumulative error in the recombination calculation may be substantial.

These errors can be much more substantial when calculating the 21-cm signal. The averaged value of the 21-cm signal in a cell is
\begin{equation}
 \langle \delta T_\mathrm{b} \rangle = \delta \hat{T}_\mathrm{b}
  \left\langle x(\mathrm{HI})(1+\delta)\left(1-\frac{T_\mathrm{CMB}}{T_\mathrm{S}}\right)\right\rangle\,;
\end{equation}
whereas, the finite volume values for the gridded quantities give the estimate
\begin{equation}
  \delta T^\prime_\mathrm{b} = \delta \hat{T}_\mathrm{b}
  \langle x(\mathrm{HI})\rangle \langle (1+\delta)\rangle \left(1-\frac{T_\mathrm{CMB}}{\langle T_\mathrm{S}\rangle}\right)\,.
\end{equation}
If we assume that $T_\mathrm{S}=T$, a fraction $f$ a cell is fully ionized and hot ($T=T^\mathrm{hot}$), and the remainder neutral and cold ($T=T^\mathrm{cold} < T_\mathrm{CMB}$), the average 21-cm signal will be
\begin{equation}
  \langle \delta T_\mathrm{b} \rangle = \delta \hat{T}_\mathrm{b}
  (1-f)(1+\delta)\left(1-\frac{T_\mathrm{CMB}}{T^\mathrm{cold}}\right) < 0.
\end{equation}
However, the quantity $\delta T^\prime_\mathrm{b}$, based on the cell averages, will be
\begin{equation}
  \delta T^\prime_\mathrm{b} = \delta \hat{T}_\mathrm{b}
  (1-f)(1+\delta)\left(1-\frac{T_\mathrm{CMB}}
  {fT^\mathrm{hot}+(1-f)T^\mathrm{cold}}\right)
\end{equation}
and positive if ${fT^\mathrm{hot}+(1-f)T^\mathrm{cold}}>T_\mathrm{CMB}$, which for $z\sim15$ and for $T^\mathrm{hot}\sim 10^4$~K is true for any $f>4\times 10^{-3}$. Clearly, we need to separate the hot and cold states to obtain the correct 21-cm signal.

Such sharp transitions in the ionization fraction and temperature are associated with relatively soft (i.e., low-energy) ionizing photons for which the mean free path is shortest. If the sources can produce both soft and hard (high-energy) ionizing photons, as is the case in the simulations presented in this paper, we need to separate their effects to establish what fraction of a cell is fully ionized by soft ionizing photons and what fraction of the cell is cold and neutral, or is partially ionized by hard photons. In Paper I, we achieved this separation by running two simulations, one with only soft sources and one with both soft and hard sources included. The difference between the two results would then give us the heating and ionization caused by the hard sources alone.

However, while this worked well for the case considered in Paper I, this solution neglects the fact that the code uses the average temperatures and electron fractions to calculate recombination and cooling rates, which are also strongly non-linear functions of the temperature. In addition, it is rather wasteful to run two simulations for every case. We therefore have introduced a different approach in which we let the code internally and self-consistently take into account the multiphase character of the ionization front cells.

This new multiphase code separates and individually tracks the effects of the photoionization and heating rates from the soft and hard sources, $\Gamma^\mathrm{soft}$ and $\Gamma^\mathrm{hard}$, and also keeps track of the mass fraction $f$ of a cell that has been ionized by soft sources. Since soft photons have very short mean free paths in neutral or partially ionized media due to their high interaction cross-sections, there is always a sharp boundary outlining the volume affected by them. Cells can be in one of three categories:
\begin{enumerate}
\item Pre-multiphase cells: these cells have never seen a soft ionizing rate above certain minimum value $\Gamma^\mathrm{soft}>\Gamma^\mathrm{soft,lim}$ and are considered to not be affected by soft ionizing photons. They have a uniform ionization fraction and temperature, determined by $\Gamma^\mathrm{hard}$. These cells produce a 21-cm signal, which is calculated based on the cell-averaged quantities.
\item Multiphase cells: these cells have at some point experienced a soft ionization rate $\Gamma^\mathrm{soft}>\Gamma^\mathrm{soft,lim}$ and a fraction $f$ of these cells affected by soft radiation is assumed to be fully ionized ($x(\mathrm{HI})=0$) and heated to $T^\mathrm{soft}$. We take $T^\mathrm{soft}=10^4$~K. The value of $f$ depends on the evolution of $\Gamma^\mathrm{soft}$. The ionization and thermal state of the rest of the cell, $1-f$, is calculated based on the hard-photon rates $\Gamma^\mathrm{hard}$. Only this $1-f$ cell fraction contains any neutral gas, and thus only it produces a 21-cm signal.
\item Post-multiphase cells: these cells have become fully ionized by soft ionizing sources, or in other words $f\approx 1$. In practice, we set a limit of $f^\mathrm{lim}=0.999$ above which we consider a cell to have reached this state. We calculate the ionization fraction and temperature based on the total rates $\Gamma^\mathrm{soft}+\Gamma^\mathrm{hard}$. These cells do not produce any appreciable 21-cm signal.
\end{enumerate}

The minimal ionization rate needed for cells to become multiphase is set to $\Gamma^\mathrm{soft,lim}=10^{-5}/\Delta t$, where $\Delta t$ is the time step. This threshold implies that the soft ionizing photons can ionize a fraction $f=10^{-5}$ of the cell within one time step. We found in tests that for the resolution and time step we use ($\sim 1$~Mpc and $\sim 5$~Myr), this limit correctly identifies ionization front location cells.

For the multiphase cells, we only calculate the hydrogen ionization fraction, which we equate to the fraction $f$ that has $x(\mathrm{HII})=1$. We assume that inside this fraction $f$ of the cell affected by soft photons the helium ionization state [$x(\mathrm{HeI}),x(\mathrm{HeII}),x(\mathrm{HeIII})$] will be $[0,1-x(\mathrm{HeIII}),x(\mathrm{HeIII})]$, where $x(\mathrm{HeIII})$ is set by the hard sources and is assumed to have this value over the entire cell.

Algorithmically, the code first considers the effect of the soft ionizing sources. By ray tracing from these sources and calculating the ionization rates from them, we establish which cells are in which of the above three categories. For the multiphase cells, we find the value of $f$. Since we assume that HeII follows HII and that the temperature is $T^\mathrm{soft}$, we only calculate the hydrogen ionization rates during this step. Because we have multiple soft sources contributing to a cell, we iterate to obtain converged values for $f$. The recombination rates are calculated assuming the temperature $T^\mathrm{soft}$.

After this step, we ray trace again, but now considering both hard and soft sources, calculating the H and He photo-ionization rates as well as the heating rates. For the pre- and post-multiphase cells, we apply the sum of the contributions of all sources. For the multiphase cells, we disregard the contribution of the soft sources, since their contribution to the ionization was accounted for in the first step. 

We therefore only apply the rates from the hard sources to the fraction $1-f$ of the multiphase cells. We once again iterate over all sources to obtain convergence on the ionization fractions and temperatures of the pre- and post-multiphase cells, as well as in the fraction $1-f$ of the multiphase cells. For the multiphase cells, we use  recombination and cooling rates calculated from the temperature of the $1-f$ fraction of the cells that follows from the heating by hard sources.

\section{Description of test simulation results}

In order to test and verify our new method, we have run a series of idealized test cases, as summarized below. First, we briefly recap the original method used to correct our simulations. As described in Paper I, the temperature calculated by the original method is given by:
\begin{equation}
T_\mathrm{HI,x} = \frac{T_\mathrm{c,x} - T_\mathrm{HII,s} x}{1-x},
\label{eq:thi}
\end{equation}
where 
$T_\mathrm{c,x}$ is the temperature from the \textsc{\small C$^2$-Ray} X-ray simulation, $x$ is the ionized fraction, and $T_\mathrm{HII,s}$ is the temperature of the ionized region given by:
\begin{equation}
T_\mathrm{HII,s} = \frac{T_\mathrm{c,s} - T_\mathrm{ad} (1-x)}{x}.
\label{eq:thii}
\end{equation}
Here, $T_\mathrm{c,s}$ is the temperature from the stellar-only run, and $T_\mathrm{ad}$ is the adiabatic temperature of the universe.

\subsection{Test 1: Comparison to a High-Resolution box}

\begin{figure} 
\centering
\includegraphics[width=1.0\columnwidth]{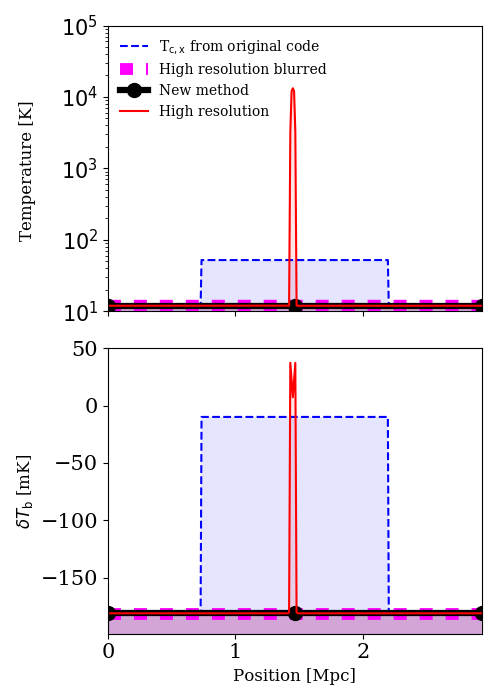}
\caption{Test 1: Comparison of our new multiphase radiative transfer method to the previous, old \textsc{\small C$^2$-Ray} code, as well as to a high-resolution simulation using the old method for a typical stellar-only source at $z=22.67$. The plot shows cross section of the kinetic temperature (top) and $\delta T_{\rm b}$ (bottom). The temperature is over estimated in the original method when compared to the high-resolution case. The multiphase method, however, yields the same result as the high-resolution run when it is smoothed to the resolution of the simulation. In the lower panel, we can see that the old method predicts a signal of emission; whereas, the new method and smoothed, high-resolution run give the expected result for a source with no X-rays, i.e. absorption from the cold gas in the surrounding cells.}
\label{fig:test1}
\end{figure}

\begin{figure*}
\centering
\includegraphics[width=1.0\linewidth]{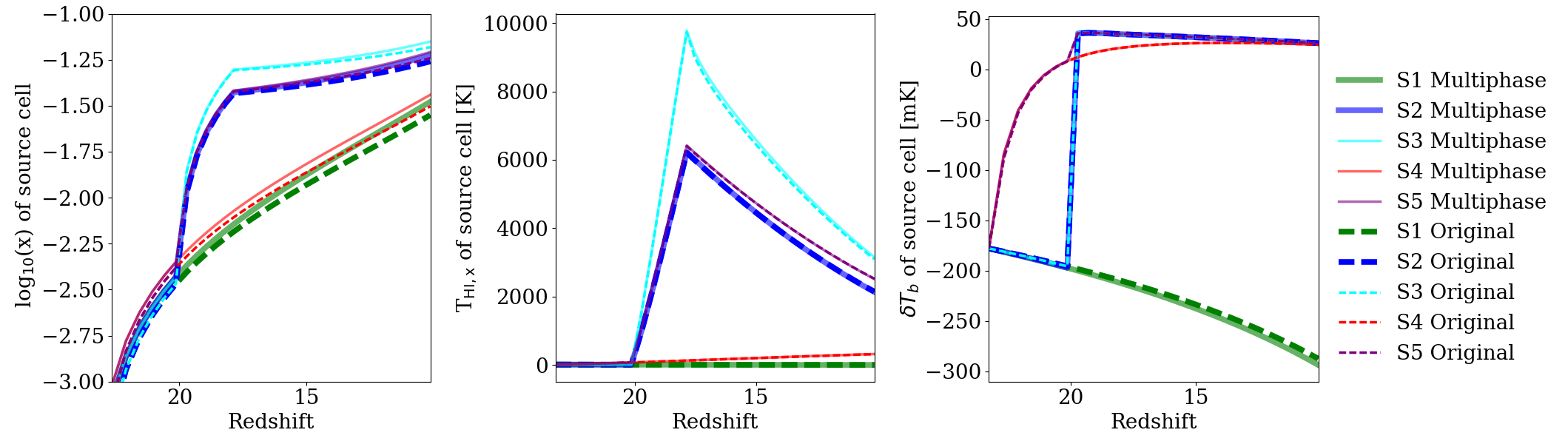}
\caption{Test results comparing the new multiphase method and the old code. On the left, we show the ionized fraction evolution with the `classic' method (dashed lines) and the new multiphase method (solid lines). In the middle panel, we show the mean temperature evolution using the multiphase method (solid lines), the `classic' method (dotted lines), and the original method after the correction described in Paper I (dashed lines). Finally, on the right, we show $\delta T_{\rm b}$, following the same notation as the middle panel.} 
\label{fig:test2}
\end{figure*}

Firstly, we compare the raw outputs of the multiphase and old version of the code both at high (3.25~kpc~$h^{-1}$ per cell) and simulation resolution (0.976~Mpc~$h^{-1}$ per cell) for a 2.928~Mpc~$h^{-1}$ box. Boxes have a single, stellar-only source in the box centre with the average luminosity of a stellar source in our simulations ($10^{50}$ ionizing photons per second), and the test is run from $z=23.268$ until $z=22.67$. 

In Fig.~\ref{fig:test1}, we show the result from Test 1 for $T_{\rm c,x}$ (top) and $\delta T_{\rm b}$ (bottom) cross-sections through the forming HII region.  The results from the high-resolution box (thin red line) are correct despite using the original method, since the ionization front has been fairly well resolved. Currently, computational resources are insufficient to run a full simulation box at this resolution; otherwise, high-resolution runs would be a valid solution.

In order to compare the results at the resolution of our large-scale simulations, we coarsen these results to the relevant resolution (thick orange line). As shown in the upper panel, the original method (dashed blue line) clearly overestimates the temperature of the neutral IGM when compared to the smoothed high-resolution run. However, the multiphase method (dotted purple line) is in agreement with the smoothed high-resolution box. 

Fig.~\ref{fig:test1} (lower panel) shows the impact this overestimation of the temperature of the neutral IGM has on $\delta T_{\rm b}$. The old code (blue dashed line) predicts $\delta T_{\rm b}$ to be much higher, and partly in emission, than the value calculated from the high-resolution box that is smoothed to simulation resolution (thick orange line). The multiphase method (dotted purple line) again agrees with the smoothed high-resolution box (thick orange line). Note that the high-resolution run does in fact show a small amount of emission due to photons from the harder end of the blackbody spectrum and (still) insufficient resolution to fully resolve the I-front. However, clearly these heated regions do not contribute significantly to $\delta T_{\rm b}$.

\subsection{Test 2: Comparison to $\boldmath{\delta T_{\rm b}}$ calculation method used in Paper I}

We compare the new multiphase method to the correction method we used in Paper I by performing two additional constant density test simulations at a resolution twice the critical density of the universe (the effect of varying the density of the boxes was checked and concluded to be minimal). 
All boxes have luminous sources in the centre with the same parameters as in Table~\ref{tab:runs}: only stars (S1), and in combination with HMXBs (S4), QSOs (S2 and S3), and all types (S5). The luminosity of HMXBs were typical of our simulations ($10^{49}$ ionizing photons per second). QSO sources have a luminosity of $10^{52}$ ionizing photons per second and do not switch on until $z\!\sim\!20$. As in the simulation, they are are active for 34.5~Myr. 

The results of these tests are shown in Fig.~\ref{fig:test2} alongside results from the lower resolution boxes considered in Test 1. As the multiphase and original method treat the cells without stellar radiation equivalently, we focus on the values of the source cell only (in both of these tests the ionized bubble never leaves this cell). In Fig.~\ref{fig:test2} (left), we show the hydrogen ionized fraction evolution produced by the two methods. The ionized fraction of hydrogen is slightly lower in the original case for all test boxes, which is due to the fact that recombination rates were previously calculated from the average temperature and hydrogen ionized fraction of the cell and therefore were systematically overestimated. In the new method, recombination rates are calculated for ionized hydrogen at the assumed temperature of the ionized region ($10^4$~K). Apart from this minor difference, both original and multiphase methods display the same smooth evolution.

In Fig.~\ref{fig:test2} (middle), we show the evolution of the $T_{\rm HI,x}$ of the source cell, comparing the new multiphase method (solid) and the original method (calculated as described in Paper I). We can see that, for all sources, the two methods are in agreement. In the right panel of this figure, we show $\delta T_{\rm b}$ for both methods and all source types. Again, the two methods are in excellent agreement.

We therefore conclude that the multiphase method and our previous correction method of calculating $\delta T_{\rm b}$ using post-processing can both be used for all sources. However, there are limitations of the original method, which are described in the following section.

\subsection{Test 3: Limitations of the Original Correction Method}
\label{sec:limits}

The multiphase and original correction methods diverge from one another when the temperature rises above 10,000~K. As discussed above, the original correction method requires two simulations: one with X-rays and one without. At temperatures greater than 10,000~K, radiative hydrogen-line cooling due to collisional excitations becomes efficient. This cooling effectively caps the gas temperatures given by the two original simulations: $T_{\rm c,s}$ and $T_{\rm c,x}$, and therefore the value for $T_{\rm HII,x}$ (calculated using Equations~\ref{eq:thii} and~\ref{eq:thi}) is underestimated and becomes unreliable. In addition, as the X-ray simulation yields slightly higher temperatures, cooling occurs more rapidly when sources are switched off. In this case, $T_\mathrm{c,s}$ can even become larger than $T_\mathrm{c,x}$, meaning that the original method can even formally yield negative temperatures in some cells.

We demonstrate this effect using a simulation-resolution test box containing a bright star (1 $\times 10^{52}$ ionizing photons per second) and a QSO (with 1 $\times 10^{51}$ ionizing photons per second) inside. The QSO switches on at $z\sim20$ and then turns off at $z\sim18$. 

\begin{figure}
\centering
\includegraphics[width=0.8\columnwidth]{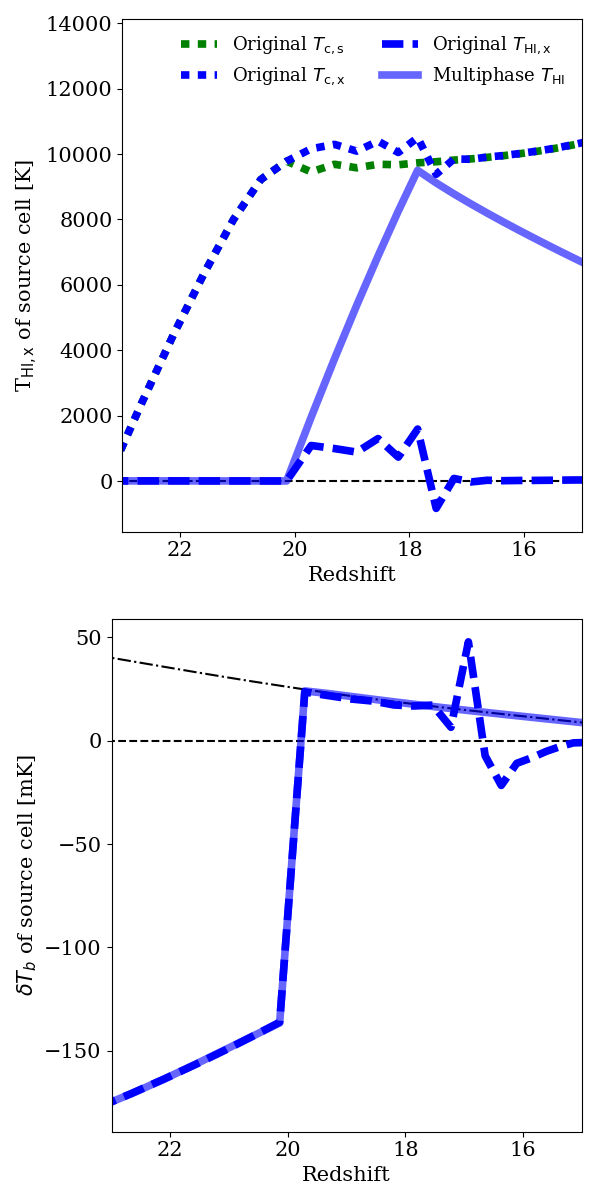}
\caption{Test runs showing the limitations of the old method. Shown are the evolution of the kinetic temperature, $T_{\rm K}$ (top), and of the differential brightness temperature, $\delta T_{\rm b}$ (bottom), for all
test simulations, as labeled. The high-$T_{\rm S}$ limit is also shown (black dot-dash line).}
\label{fig:test4}
\end{figure}

The top panel of Fig.~\ref{fig:test4} shows the temperature of the source cell. The value of $T_{\rm HI}$ from the new multiphase version of the code is plotted using the solid blue line. This temperature remains low at early times as there is no X-ray heating. When the QSO switches on, $T_{\rm HI}$ from the multiphase version of the code continues to rise until the QSO switches off. After this, $T_{\rm HI}$ cools slowly, remaining well above $T_{\rm CMB}$. 

The values of T$_{\rm c,s}$ (green dotted line) and $T_{\rm c,x}$ (blue dotted lines) from the original code are also shown in the top panel of Fig.~\ref{fig:test4}. Both agree until the QSO becomes active, at which point $T_{\rm c,x}$ rises above $T_{\rm c,s}$ due to X-ray heating, as expected. When the QSO switches off, $T_{\rm c,x}$ is greater than T$_{\rm c,s}$, so the X-ray simulation cools more rapidly than the stellar simulation and $T_{\rm c,x}$ briefly drops below T$_{\rm c,s}$. After this, the star in the cell continues heating and ionizing and drives the T$_{\rm c,s}$ and T$_{\rm c,x}$ back up to 10,000~K. 

\begin{figure}
\centering
\includegraphics[width=0.8\columnwidth]{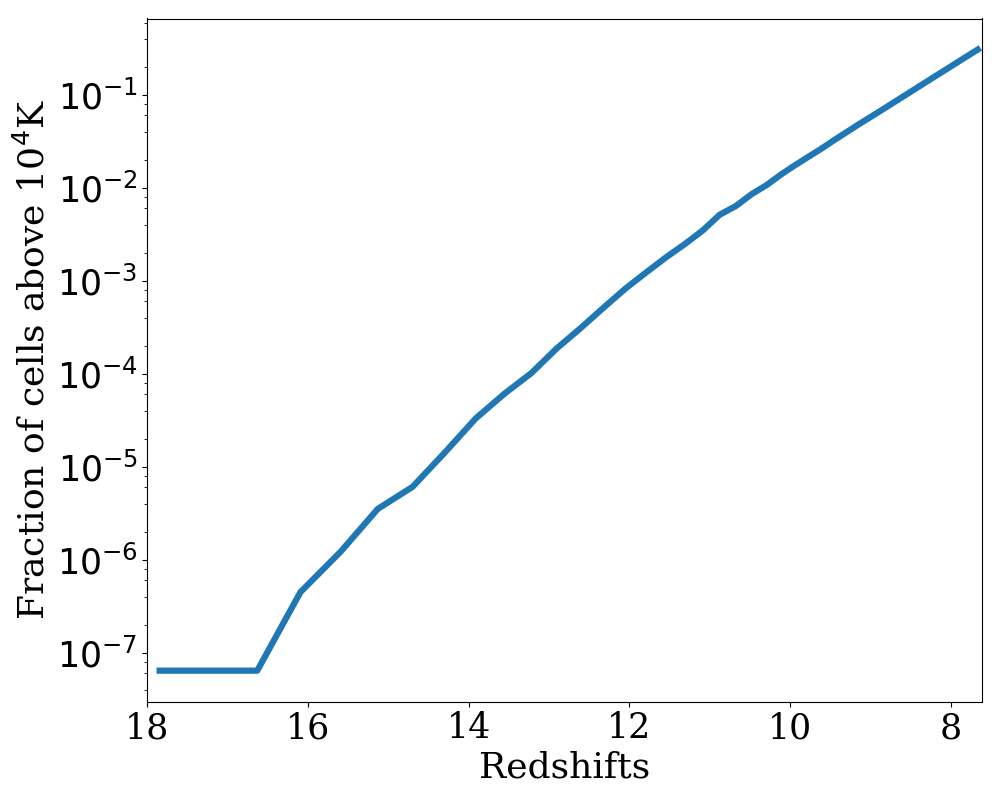}
\caption{Shown are the fraction of cells that exceed 10,000~K in the stellar-only simulation for the original code. Before $z\approx 18$ there are no cells above 10,000~K. The number remains small before $z\approx 11$ showing that the original corrections method can be used on cells with HMXBs which reach temperature saturation at $z>12$.}
\label{fig:wrong}
\end{figure}

The value of $T_{\rm HI,x}$ calculated from T$_{\rm c,s}$ and T$_{\rm c,x}$ (as described in Equations~\ref{eq:thi} and \ref{eq:thii}) is shown with the dashed blue line. The temperature is in agreement with the value given by the mutliphase method as long as there is no X-ray heating. When the QSO switches on, $T_{\rm HI,x}$ briefly increases before leveling. This plateau happens as $T_{\rm c,s}$ and $T_{\rm c,x}$ are capped by collisional cooling, meaning that the X-ray heating is vastly underestimated. When the QSO switches off, $T_{\rm HI,x}$ drops below zero (zero is marked with the black dashed line to illustrate where this occurs). This behavior is clearly unphysical and is due to T$_{\rm c,x}$ dropping below T$_{\rm c,s}$. After this, $T_{\rm HI,x}$ returns to its initial, low value as both $T_{\rm c,s}$ and $T_{\rm c,x}$ are capped at approximately 10,000~K. Hence, X-ray heating is underestimated. 

In the lower panel of Fig.~\ref{fig:test4}, $\delta T_{\rm b}$ is shown. $\delta T_{\rm b}$ in both methods increases before there is X-ray heating due to the expanding ionized region created by the bright star in this case, which gradually decreases the neutral fraction. As described above, the multiphase and original correction method agree until until the QSO switches on at $z \approx 20$. When X-ray heating from the QSO begins, $\delta T_{\rm b}$ increases rapidly until temperature saturation is reached (the high-$T_{\rm S}$ limit is marked on with the black dot-dash line). As the $T_{\rm HI}$ from the multiphase method remains above $T_{\rm CMB}$ for the rest of the simulation, $\delta T_{\rm b}$ follows the high-$T_{\rm S}$ limit and slowly decreases towards zero as the ionized region from the star continues to grow. $\delta T_{\rm b}$ from the original corrections method agrees with the multiphase value until the QSO switches off. At this point, $T_{\rm HI,x}$ drops rapidly and becomes unphysical, which causes erratic behavior in the $\delta T_{\rm b}$ values obtained using the original method. 

This effect does not impact simulations including HMXBs, since in these simulations temperature saturation is reached at $z\sim12$, which is before any non-negligible number of cells are heated above 10,000~K (see Fig.~\ref{fig:wrong}). We confirmed that the results from S4 and S5 (run with the original method) are the same as those given by the multiphase method by running parts of the simulation
with both code versions.

\label{lastpage}
\end{document}